\documentclass[11pt]{article}
\usepackage[a4paper,margin=2.5cm]{geometry}
\usepackage[T1]{fontenc}
\usepackage[utf8]{inputenc}
\usepackage{amsmath,amssymb}
\usepackage{graphicx}
\usepackage{booktabs}
\usepackage[numbers]{natbib}
\usepackage{tikz}
\usetikzlibrary{arrows.meta}
\usepackage[british]{babel}
\PassOptionsToPackage{hyphens}{url}
\usepackage[hidelinks]{hyperref}
\graphicspath{{figure/}}

\title{Scale, Concentration, and Entry Timing in the Shopify App
Ecosystem: A Longitudinal Study of Platform Governance and
Application Survival}

\author{Fabrizio Assabese\\Judge.me LTD, London, United Kingdom\\
\texttt{fabrizio@judge.me}
\and
Peter-Jan Celis\\Judge.me LTD, London, United Kingdom\\
\texttt{pj.celis@judge.me}
\and
Giuseppe Destefanis\\Department of Computer Science, UCL, London, United Kingdom\\
\texttt{g.destefanis@ucl.ac.uk}}

\date{}

\begin{document}
\maketitle

\begin{abstract}
The Shopify marketplace hosts more than 16,000 active third-party
applications, serving 2.7 million active merchant stores generating an estimated \$706 billion in annual sales, yet little empirical evidence exists
on its structure and adoption drivers. We analyse it using a
September 2025 snapshot of all 24,826 applications, a weekly panel of
7,708 applications over 366 weeks (February 2019 to March 2026), and
listing histories reconstructed from the Internet Archive. Across 55
established functional categories, over half were low-concentration
and 20\% highly concentrated, with larger categories consistently
less concentrated. Later entrants grew faster than early movers in 88\% of the 50 analysed categories, persisting across seven years of quarterly re-estimations, with early movers on net losing detected installations whilst late movers gained them. Platform governance reshaped
competition asymmetrically: Shopify's entry into chat reversed more
than two years of de-concentration; the deprecation of its Product Reviews application, to our knowledge the first measured platform-owner exit from a complementor category, released 191,000 installations of which at
most a third reappeared as competitor adoption within a year; the
2021 reduction of the platform revenue share produced no detectable
change in entry or retention. Failure is largely silent and
predictable: the median exiting application peaked at 8 detected installations and disappeared from tracking within 68 weeks of launch, category leadership
changed hands in 92\% of categories over seven years, and publicly
observable data from an application's first six months predict exit
within two years with cross-validated AUC above 0.8. App markets on
Shopify remain contestable; who benefits depends on platform
governance and entry conditions more than on entry timing.
\end{abstract}

\vspace{0.5em}

\section{Introduction}\label{sec:introduction}

Shopify\footnote{\url{https://www.shopify.com/}} is a platform that enables businesses to create and operate online stores. It provides merchants with the technical foundation to sell products, handle payments, and manage orders, with subscription plans ranging from small independent shops to large international brands. A distinctive aspect of Shopify's model is that many advanced functions are not built into the core product but supplied through an external marketplace of applications. These applications, developed by third-party companies, extend store capabilities with features such as product reviews, marketing automation, logistics, subscription management, and integration with social media or advertising platforms.

This marketplace now comprises more than 16{,}000 active applications maintained by thousands of independent software-producing organisations. For merchants, it is central to how storefronts are designed and how they compete in online retail. For developers, it offers access to a global market of potential customers. Several applications that began as small projects have grown into significant companies, showing that the marketplace operates both as infrastructure for commerce and as a channel for software entrepreneurship.
Developers considering whether to build for Shopify need to know which categories are saturated, whether early entrants retain growth advantages, and how competition evolves. Merchants must evaluate the risks of relying on dominant providers in essential categories whilst assessing whether newer alternatives provide genuine improvements. For Shopify itself, adoption patterns across applications affect openness, competition, and long-term sustainability. These questions can now be studied systematically by combining adoption data from Store Leads\footnote{\url{https://storeleads.app/}} with listing histories reconstructed from the Internet Archive.\footnote{\url{https://archive.org/}}

Store Leads scrapes Shopify storefronts on a weekly basis, providing datasets with information on both domains (including categories, estimated revenue, and installed applications) and applications (including installs, creation date, ratings, and vendor details). Because the Shopify App Store displays only current listings, we complement this adoption data with a second source built for this work: a historical record of listing characteristics (pricing, categories, vendor, reviews) that we reconstructed from archived copies of App Store pages in the Internet Archive's Wayback Machine (Section~\ref{sec:dataset}). Together these sources make it possible to analyse adoption dynamics, market concentration, and entry timing at scale. In this study, we use them to investigate how the Shopify marketplace operates as an ecosystem and what this reveals about opportunities and constraints for software-producing organisations.

This article is an extended version of the IWSiB '26 paper by the first and third authors
\citep{assabese2026scale}. We retain the original study's three
research questions and their answers, and build the longitudinal
analysis on them:\\
\noindent
\textbf{RQ1:} What are the dynamics and scale of the Shopify app ecosystem? \textit{Rationale:} Establishing the size of the ecosystem, its temporal evolution, the distribution of applications across categories, geographic concentration, and adoption patterns provides the context for subsequent analyses. Understanding ecosystem growth and merchant deployment patterns frames the competitive environment that developers enter and clarifies whether building for Shopify represents a viable business direction.\\
\noindent
\textbf{RQ2:} How concentrated is the app ecosystem within specific categories? \textit{Rationale:} Categories vary widely in scale and importance. Some attract intense developer investment and competition, while others are dominated by only a few providers. Assessing whether adoption is concentrated or distributed is essential for evaluating competitive dynamics and identifying where opportunities for new applications remain.\\
\noindent
\textbf{RQ3:} Do early entrants maintain growth advantages over time? \textit{Rationale:} Although cumulative installation counts favour older applications through longer exposure, recent growth patterns reveal whether early movers sustain momentum or lose ground to newer entrants. By analysing both absolute and proportional growth, we test whether entry timing confers lasting advantage or whether markets remain contestable as categories mature.

The original study \citep{assabese2026scale} answered these questions
with a snapshot collected in September 2025, finding substantial
heterogeneity in competitive structure, an inverse relationship
between category size and concentration, and a systematic late-mover
advantage: in 88\% of the 50 analysed categories, later entrants grew faster over the preceding 90 days, whilst many early movers lost detected installations on net.

A single snapshot, however, can only measure the market once. All
growth quantities in the original study derive from one 90-day window,
so the central finding could in principle reflect conditions specific
to mid-2025; and questions about how competition itself evolves, what
happens when the platform owner intervenes in a category, and which
entrants survive cannot be asked cross-sectionally at all. This
article extends the original study with a weekly panel of 7,708
applications observed over 366 weeks, from February 2019 to March
2026, linked to the original snapshot and to listing characteristics
reconstructed from the Internet Archive at semi-annual resolution.
The panel places the original results in time and enables us to study two questions that a single snapshot cannot address: platform governance and application survival.

The first is governance. During the panel period Shopify entered
complementor categories (Shopify Inbox in chat, Shopify Forms in email
capture), left one (the deprecation of its Product Reviews
application), and changed developer economics (the reduction of its
revenue share to 0\% below \$1M). Each event is visible at weekly
resolution in the adoption data, allowing us to measure how
marketplace orchestration reshapes complementor competition. To our
knowledge the deprecation provides the first measured case of platform-owner exit from a complementor category, complementing the extensive literature on platform-owner entry.

The second is survival. A snapshot describes only the applications
present when it is taken, so it cannot show how much failure the
marketplace produces, how quickly failing applications disappear, or
whether that failure is foreseeable from observable signals. Observing
exits directly at weekly resolution turns the original account of contestability into one that includes application exits.

We therefore add three research questions to the original three:\\
\noindent
\textbf{RQ4:} How do platform governance events reshape competition
within affected categories? \textit{Rationale:} During the panel
period Shopify entered complementor categories, exited one, and
changed developer economics. Prior work has studied platform owner
entry on other platforms, but to our knowledge exit has not been measured, and commission changes are rarely observable alongside adoption data.
Establishing how each intervention alters incumbent growth and
category concentration shows which governance instruments actually
move competitive structure, informing both developers assessing
platform risk and operators weighing intervention.\\
\noindent
\textbf{RQ5:} What predicts application survival, and what does the
typical application lifecycle look like? \textit{Rationale:}
Cross-sectional growth comparisons only describe applications that are
still present. Observing exits directly reveals how much failure the
marketplace produces, how quickly it occurs, and which observable
conditions at entry (pricing model, category concentration, early
traction) are associated with lasting presence,
turning the original study's account of contestability into a full picture that includes application exits.\\
\noindent
\textbf{RQ6:} Can publicly observable data from an application's first
months predict its survival? \textit{Rationale:} If early adoption
signals carry most of the information about eventual exit, developers can recognise failing positioning while repositioning remains feasible, and
platform operators and merchants can assess the durability of an
application from observable data. A cross-validated predictive model
also quantifies how much of marketplace failure is foreseeable rather
than random, complementing the explanatory models of RQ5 with an
automated decision-support instrument.

\paragraph{Relation to the original paper.}
This article extends \citet{assabese2026scale}. Retained from the
original are the motivation, the related work, the snapshot dataset
description, the measurement apparatus (market shares, HHI and its
thresholds, the velocity and rate decomposition), and the full RQ1--RQ3
results (Sections~\ref{sec:rq1}--\ref{sec:rq3}). The following
material is new:
\begin{itemize}
\item two longitudinal data sources and their linkage to the
  snapshot: the seven-year weekly installation panel and the listing
  histories reconstructed from the Internet Archive
  (Section~\ref{sec:dataset});
\item a verification of the original results: we recomputed every
  headline number from the raw snapshot with newly written code,
  checked that the panel and the snapshot agree where they overlap,
  and re-estimated the entry-timing result at 26 quarterly points
  across seven years to confirm it is not specific to one measurement
  window (Section~\ref{sec:rq123});
\item the platform governance event studies of RQ4
  (Section~\ref{sec:rq4});
\item the survival and lifecycle analysis of RQ5
  (Section~\ref{sec:rq5});
\item the early-warning prediction model of RQ6
  (Section~\ref{sec:rq6});
\item robustness checks that validate the exit measure against
  snapshot status, test for tracking-selection effects, and compare
  each event-study contrast against all other categories over the
  same period;
\item the discussion (Section~\ref{sec:discussion}) and a replication
  package, whose contents and data-licensing limits
  are set out in the Data availability statement.
\end{itemize}

\section{Related Works}\label{sec:related}

Platform ecosystems depend on complementors who develop applications on top of the platform infrastructure \cite{ceccagnoli2012cocreation}. Early empirical work on mobile app stores documented rapid complementor proliferation and innovation patterns \cite{boudreau2012let}, establishing foundational questions about openness, governance, and value appropriation in two-sided markets. Subsequent research has examined complementor dynamics through three primary lenses: platform owner entry effects, governance mechanisms, and market structure.

A body of work analyses how platform owner entry into complementor spaces affects competition and innovation. Studying Android, Wen and Zhu \cite{wen2019threat} found that threats of Google entry reduced complementor innovation whilst raising prices, whereas Foerderer et al. \cite{foerderer2018does} documented attention spillovers from Google Photos entry that increased innovation in affected categories. Similar dynamics emerge in retail marketplaces, where Amazon's entry into third-party product categories reduces small-seller growth \cite{zhu2018competing} and triggers offline disintermediation strategies \cite{he2020impact}. Platform governance through awards \cite{foerderer2021and}, endorsements \cite{agarwal2023growing}, and venture capital signals \cite{van2023anchored} shapes complementor entry decisions and resource allocation. Whilst these studies establish how platforms influence complementor behaviour, they do not systematically measure adoption distributions or concentration across functional categories.

Research on market structure in platform ecosystems remains limited. One longitudinal study of mobile app usage documented Pareto-like distributions and intra-category competitive elimination \cite{li2021understanding}, but systematic category-level concentration measurement using formal metrics across revenue or install bases is scarce. Quantitative ecosystem evolution studies exist for open-source platforms like R \cite{plakidas2016software}, but comparable large-scale analyses of commercial B2B marketplaces remain absent. Studies of generativity tension in video game console ecosystems \cite{cennamo2019generativity} and cross-platform responses to entry \cite{kapacinskaite2024competing} examine developer strategies but not aggregate market structure.

Enterprise and B2B platform ecosystems have received less empirical attention than consumer app stores. Prior work has examined complementor entry decisions in enterprise software platforms \cite{huang2009isvs}, venture capital signalling effects in Salesforce \cite{van2023anchored}, and innovation responses to antitrust intervention in Microsoft ecosystems \cite{thatchenkery2023innovation}. However, large-scale quantitative studies of B2B marketplace structure and post-entry growth dynamics, comparable to iOS/Android research, remain absent.

Our study provides a platform-wide empirical analysis of the Shopify ecosystem, a major B2B e-commerce platform serving 2.7 million merchant stores. We contribute systematic category-level concentration measurement across 55 functional categories using HHI and top-5 market share metrics, document substantial heterogeneity in competitive structure (challenging universal winner-take-all assumptions), and establish that late entrants systematically outperform early movers in recent growth across 88\% of the 50 analysed categories, with early movers on net losing detected installations whilst late movers capture growth. We focus on supply-side market structure in a B2B context, providing quantitative evidence on concentration dynamics and entry timing effects in an underexamined but economically significant platform ecosystem.

\section{Dataset}\label{sec:dataset}

The study combines three sources: the cross-sectional snapshot used
in the original paper, a weekly installation panel from Store Leads,
and a historical record of listing characteristics that we
reconstructed from archived copies of App Store listing pages. The
panel provides the outcome variable, detected installations, at
weekly frequency; the reconstruction provides time-varying covariates
(pricing, categories, vendor, reviews) that are not otherwise
available historically, because the Shopify App Store displays only
current listings.

We use data from Store Leads\footnote{\url{https://storeleads.app/}}, a commercial provider that performs weekly crawls of Shopify storefronts and the Shopify app store. Our snapshot dataset, collected on 28 September 2025, consists of two files: an application-level export containing 24{,}826 Shopify apps and a domain-level export containing 2{,}701{,}805 Shopify stores.\footnote{The application-level export of 28 September is retained and reproduces every application-side figure in Section~\ref{sec:rq1}. For the domain-level data the retained file is the pull of 20 September 2025, holding 2,684,379 stores; the store-side figures in Section~\ref{sec:rq1} derive from the 28 September pull of the original study, and the two vintages agree to within 1\% on the headline quantities (total estimated sales \$704.1 billion against \$706.3 billion; United States share 38.5\% against 38.3\%). The largest divergence is the India store count, 4.9\% lower in the archived pull, which also swaps India's rank with Australia's.} The store count covers live merchant stores with an active storefront; development stores and closed stores are not included, so the total number of Shopify websites is considerably higher.

The application file includes: unique identifier, name, creation date, current status (active or removed), functional categories (colon-separated labels such as product reviews, marketing, SEO), pricing (minimum and maximum monthly subscription in USD), adoption metrics (total installs, installs in the last 30 and 90 days), and quality signals (average rating, total reviews, recent review counts). Vendor fields cover name, email, website, and registered address. The domain file includes merchant name, store creation date, geographic location (city, country code), subscription plan, estimated yearly sales in USD (derived from traffic models), average product price, installed app count, and a colon-separated list of installed app names.

Install counts represent the cumulative number of stores that installed each application, as detected by Store Leads through storefront scraping. Merchants typically remove inactive application code from their storefronts to maintain site performance, suggesting that detected installations predominantly represent actively used apps rather than abandoned integrations. Store Leads tracks both cumulative installations and recent changes (installations in the last 30 and 90 days), with the latter capturing net changes including removals. Our analyses use cumulative installations to measure total market penetration (RQ1, RQ2) and recent installation metrics to assess growth dynamics (RQ3).

Dates were provided in string format and converted to date objects.
Application categories and store tags were standardised through case-insensitive matching.

Our analytical sample consists of active applications (status = Active) with at least one installation. This criterion excludes abandoned projects, pre-launch listings, and removed apps whilst retaining niche applications serving small merchant groups. Of the 24{,}826 applications, 16{,}698 were active, and 4{,}213 of these had at least one installation. This filtered sample forms the analytical sample for the concentration and entry-timing analyses (RQ2--RQ3); the scale description of RQ1 also draws on the full active set.

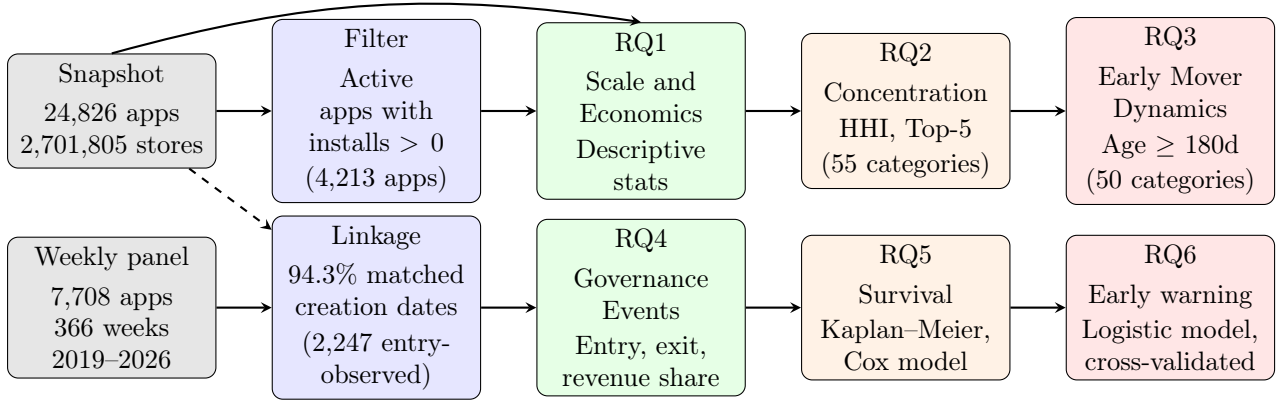
\begin{figure*}[htb!]
\centering
\begin{tikzpicture}[
  node distance=3.5cm,
  box/.style={rectangle, draw, text width=2.5cm, align=center, minimum height=1.5cm, font=\small, rounded corners},
  arrow/.style={->, >=stealth, thick}
]
\node[box, fill=gray!20] (data) {Snapshot\\[0.1cm]24,826 apps\\2,701,805 stores};
\node[box, right of=data, fill=blue!10] (filter) {Filter\\[0.1cm]Active apps with\\installs $> 0$\\[0.05cm](4,213 apps)};
\node[box, right of=filter, fill=green!10] (rq1) {RQ1\\[0.1cm]Scale and\\Economics\\[0.05cm]Descriptive stats};
\node[box, right of=rq1, fill=orange!10] (rq2) {RQ2\\[0.1cm]Concentration\\[0.05cm]HHI, Top-5\\[0.05cm](55 categories)};
\node[box, right of=rq2, fill=red!10] (rq3) {RQ3\\[0.1cm]Early Mover\\Dynamics\\[0.05cm]Age $\geq$ 180d\\[0.05cm](50 categories)};
\draw[arrow] (data) -- (filter);
\draw[arrow] (filter) -- (rq1);
\draw[arrow] (rq1) -- (rq2);
\draw[arrow] (rq2) -- (rq3);
\node[box, below of=data, node distance=2.6cm, fill=gray!20] (panel) {Weekly panel\\[0.1cm]7,708 apps\\366 weeks\\2019--2026};
\node[box, right of=panel, fill=blue!10] (link) {Linkage\\[0.1cm]94.3\% matched\\creation dates\\[0.05cm](2,247 entry-observed)};
\node[box, right of=link, fill=green!10] (rq4) {RQ4\\[0.1cm]Governance\\Events\\[0.05cm]Entry, exit,\\revenue share};
\node[box, right of=rq4, fill=orange!10] (rq5) {RQ5\\[0.1cm]Survival\\[0.05cm]Kaplan--Meier,\\Cox model};
\node[box, right of=rq5, fill=red!10] (rq6) {RQ6\\[0.1cm]Early warning\\[0.05cm]Logistic model,\\cross-validated};
\draw[arrow] (panel) -- (link);
\draw[arrow] (link) -- (rq4);
\draw[arrow] (rq4) -- (rq5);
\draw[arrow] (rq5) -- (rq6);
\draw[arrow, dashed] (data) -- (link);
\draw[arrow] (data.north) to[bend left=12] (rq1.north);
\end{tikzpicture}
\caption{Methodological workflow. Top row: snapshot analyses (RQ1--RQ3, original study). Bottom row: weekly-panel analyses (RQ4--RQ6); the dashed arrow marks the snapshot linkage that supplies creation dates and category definitions to the panel. The curved arrow from the snapshot to RQ1 marks that RQ1's scale description draws on the full and active application sets directly; the installation filter defines the analytical sample of RQ2 and RQ3.}
\label{fig:methodology}
\end{figure*}

\paragraph{Weekly installation panel (2019--2026).}
The panel comes from Store Leads, which detects application
installations by crawling live Shopify storefronts. For each
application it records the number of live stores with the application
installed, observed weekly. Our extract contains 997,713
application-week observations covering 7,708 applications\footnote{One application appears in the extract under two identifier variants (a leading-slash duplicate covering eight early weeks), so unique applications number 7,707 and one of the 1,224 tracking exits is the short-lived variant. We retain the extract's identifiers throughout; no reported percentage changes at the stated precision.} over 366 weeks, from 18 February 2019 to 1 March 2026. Because the counts are
measured by an external observer,
they form a consistent adoption measure across the ecosystem; they
capture detectable installations on active stores, so applications
with no storefront-visible footprint may be undercounted. The data
were used as provided: our only transformations were normalising
application identifiers to the URL handle, the join key used
throughout, and computing week-over-week changes. Week spacing is seven days, with one-day jitter, occasional shifted crawl anchors, and two skipped crawl weeks leaving gaps of up to sixteen days; growth quantities are normalised per day and rescaled to weekly values before analysis.

\paragraph{Historical listing characteristics (2012--2026).}
Shopify publishes no historical record of App Store listings, so we reconstructed listing histories from the Internet Archive's Wayback Machine\footnote{\url{https://web.archive.org/}} in four stages. First, discovery: we enumerated all archived
URLs under \texttt{apps.shopify.com} through the Wayback Machine's
index of captures (the CDX API), which yielded 26,709 application
handles with at least one archived listing page. Because repricing and recategorisation are infrequent,
we sampled each application's archive at semi-annual resolution,
selecting for every application-year the two snapshots closest to 1
January and 1 July; this produced a fetch queue of 145,081 snapshot
URLs spanning 2012 to 2026. Second, acquisition: archived pages were
fetched with a rate-limited, resumable crawler and stored as
compressed raw HTML in a local database, so that extraction logic
could be revised and re-run without re-crawling; 137,040 snapshots
(94.5\%) were retrieved, and the failures are predominantly snapshots
that the archive lists in its index but cannot serve. Third,
extraction: from each snapshot we extracted pricing plans, categories,
and listing metadata, with parsing rules specific to each page layout
the App Store used over the period. Pricing extraction tries
structured page data first, then the pricing markup of each template
era, then a conservative pattern-based fallback, and yielded 251,972
plan-level observations covering 22,813 applications. Category
extraction uses five era-specific rules corresponding to successive
page layouts, yielding 167,628 category observations
extracted directly from 116,872 snapshots (85.3\% of those
retrieved); with the nearest-snapshot imputation described under
panel construction, 98.5\% of retrieved snapshots carry a category,
covering 22,397 applications. A dedicated
rule keeps only the listing's own categories and discards site-wide
navigation menus, which a naive extraction would conflate with them.
Listing metadata (vendor name, cumulative review count and rating)
was extracted for 130,268 snapshots, giving a low-frequency record of
review accumulation. Fourth, validation: parsed output was checked
against rendered archived pages from every layout era, and negative
results, such as categories appearing unrecoverable for an era, were
re-tested against alternative rules before being accepted.

\paragraph{Panel construction.}
The two sources were merged on the application handle into a single
application-week panel. Listing characteristics are observed
semi-annually whilst installations are weekly, so each week is
assigned the most recent snapshot on or before it (a
last-observation-carried-forward join), and the source snapshot date
is retained alongside every covariate, so the age of the information
is explicit and analysable. Snapshots lacking a directly extracted
category (about 14\% of category rows) were imputed from the nearest
snapshot of the same application and flagged, keeping observed and
imputed values distinguishable. The archival review counts, which are
semi-annual, were supplemented for May 2024 to May 2026 with weekly review counts provided by Store Leads (covering 17,059 applications, 6,032 of which appear in the panel), matched by week with a six-day tolerance; both series
are retained, the archival series for its 2019 to 2024 depth and the
weekly series as the preferred measure where available (96.9\% of
in-window weeks matched for covered applications).

\paragraph{Linkage.}
Panel applications are linked to the snapshot through the application
handle: 7,270 of 7,708 (94.3\%) match a snapshot record, which
provides creation dates, status in September 2025, and the original
study's category definitions. At the overlapping week (28 September
2025), panel and snapshot installation counts agree exactly for the
4,939 linked applications with positive snapshot installs (rank
correlation 1.000, median relative difference 0.0\%), confirming that
the two sources reflect the same underlying measurements.

\paragraph{Coverage and censoring.}
The panel is unbalanced: 779 applications are tracked in the first
week and 6,307 in the last. First appearance in the panel therefore
reflects tracking onboarding and does not mark market entry. We use snapshot
creation dates to separate the two: 2,247 linked applications are
\emph{entry-observed} (first tracked within 90 days of creation) and
form the basis of all cohort and survival analyses, whilst 5,023 are
left-censored (created more than 90 days before tracking began).
Applications last observed more than eight weeks before the end of the
panel are treated as exits from tracking; 1,224 applications (15.9\%)
meet this condition. Compared with the full store, the panel
overrepresents established applications, so panel-based results are
conditional on the tracked population; the replication bridge in
Section~\ref{sec:rq123} quantifies the agreement where the sources
overlap.

\section{Methodology}\label{sec:method}

We analyse the Shopify app ecosystem in three stages aligned with our research questions. First, we establish ecosystem dynamics and scale using descriptive statistics (RQ1). Second, we measure market concentration within functional categories using the Herfindahl--Hirschman Index and top N market shares (RQ2). Third, we test whether early entrants retain growth advantages by relating application creation timing to recent growth metrics rather than to cumulative share (RQ3).
Figure~\ref{fig:methodology} summarises the workflow.\\
\noindent
\textbf{\textit{Market share definition.}} For RQ2 and for descriptive context in RQ1, we compute each application's market share within its primary category as
\[
s_{i,c} \,=\, \frac{\textit{installs}_i}{\sum_{j=1}^{n_c} \textit{installs}_j},
\]
where $\textit{installs}_i$ is the cumulative installation count of application $i$ in category $c$ with $n_c$ applications.

\noindent
\textbf{\textit{RQ1: Dynamics and scale.}}
We report application counts by operational status and market participation, derive a single primary category per app from the first label in \texttt{app\_store\_categories}, and describe category size distributions. We characterise the merchant side using estimated yearly sales, installed app counts per store, creation dates, geographic codes, and pricing models classified as freemium when \texttt{min\_price} equals \$0 and paid otherwise. We identify 231 primary categories among active applications and, for later stages, the subset with sufficient competition.

\noindent
\textbf{\textit{RQ2: Concentration within categories.}}
We restrict to categories with at least 20 active applications that have installs greater than zero. This focuses on established markets and yields more stable concentration estimates. For each category, we compute
\[
\text{HHI}_c \,=\, \sum_{i=1}^{n_c} s_{i,c}^{2},
\]
and classify concentration using: HHI below 0.15 low, HHI from 0.15 to below 0.25 moderate, HHI at least 0.25 high \cite{doj2010merger}.\footnote{These thresholds were calibrated for revenue shares in merger review and serve here as a descriptive banding of installation shares. The banding is threshold-sensitive: under the earlier 1992 guidelines (0.10 and 0.18 on the unit scale), 19 of the 55 categories (34.5\%) would classify as highly concentrated and 7 of the top 20 by installations would exceed the high threshold, against 11 (20.0\%) and 3 under the thresholds used here. The continuous HHI values reported throughout are unaffected by this choice.} As a complement, we report the cumulative share of the top five applications. We examine how concentration relates to category size and total installations.

\noindent
\textbf{\textit{RQ3: Early mover growth dynamics.}}
We test whether earlier entry is associated with stronger recent growth. We use the app creation date as a proxy for entry timing and analyse only applications aged at least 180 days to avoid launch volatility. Growth is decomposed into two metrics per app: velocity, the number of installs in the last 90 days, and rate, the ratio of installs in the last 90 days to cumulative installs. Within each category we rank apps by creation date, then compute Spearman correlations between creation rank and each growth metric. Negative correlations for velocity indicate earlier apps adding more installs in absolute terms. Negative correlations for rate indicate earlier apps growing faster proportionally. Positive signs indicate advantages for later entrants. Categories are labelled as Strong FMA when both correlations are negative, FMA Eroding when velocity is negative and rate is positive, Late Mover Advantage when both are positive, and Mixed or none when signs conflict or effects are near zero. Statistical significance is assessed at $p<0.05$ with Benjamini--Hochberg false discovery rate control across categories.

\noindent
\textbf{\textit{Acceleration analysis.}}
To detect shifts in momentum, we compare 30 day installs with the 90 day average divided by three. Apps with higher 30 day value are marked as accelerating. We define early movers as the first quartile by creation date and late movers as the fourth quartile. Within each category we test differences in acceleration between these groups using the Mann--Whitney U test with false discovery rate control and report Cohen's $d$ as an effect size.

\noindent
\textbf{\textit{Scenario projections.}}
To contextualise observed gaps, we project median early and late mover baselines under three simple scenarios for 20 quarters and report 24 month outcomes as a practical horizon. The scenarios are exponential growth with constant rate, linear growth with constant absolute velocity, and a decay model with rates declining linearly to zero over the horizon. We also report time to parity, defined as the number of quarters for late movers to reach 50\% of early mover installs under each scenario. These are illustrative calculations under stated assumptions rather than forecasts.

\noindent
\textbf{\textit{Inclusion criteria.}}
All inferential analyses use mature applications aged at least 180 days and categories with at least 20 active apps with installs greater than zero. The original study's entry-timing sample additionally required a computable acceleration ratio; Section~\ref{sec:rq123} documents that rule and reproduces the published sample from it. Early and late mover groups are defined as the first and fourth quartiles by creation date within each category.

\noindent
\textbf{\textit{Replication bridge.}}
Before any longitudinal analysis, we recompute the original paper's
headline results from the raw snapshot with independent code and
verify that panel and snapshot agree at their overlapping week
(Section~\ref{sec:rq123}). We also test whether the original
entry-timing result depends on its measurement window by recomputing
the velocity and rate correlations on a quarterly grid across the
whole panel, at each point using the applications and categories that
satisfy the original inclusion rules at that date. The original
90-day growth metrics were dictated by the snapshot, which only
reports installations over fixed trailing windows; the weekly panel
allows alternatives, so we repeat the grid with a 26-week window and
with velocity measured as the fitted weekly trend over every
observation in the window, to check that neither the window length
nor the use of endpoints drives the result. The acceleration analysis
is re-tested in the same way, with acceleration measured directly as
the change in velocity between two consecutive 13-week windows.

\noindent
\textbf{\textit{RQ4: Platform governance events.}}
We study four platform governance events at weekly resolution: the
relaunch of Shopify Inbox into the chat category (July 2021), the
launch of Shopify Forms (November 2022), the deprecation of Shopify's
Product Reviews application (delisted 5 September 2023, shut down 6
May 2024), and the reduction of the platform revenue share to 0\% on
the first \$1M of developer revenue (effective 1 August 2021). Event
dates come from public announcements. For category events we examine
installation trajectories of the platform application and the leading
incumbents, pre- and post-event growth over 26-week windows, weekly
category HHI, and the redistribution of installations after the
shutdown. Redistribution separates gross competitor gains from
\emph{excess} gains, defined as each competitor's post-event 52-week
growth minus its own pre-event 52-week growth, so that pre-existing
trends are not attributed to the event; the recovery share is total
positive excess divided by the platform application's loss. Event
contrasts are placed against the ecosystem with a permutation check:
the affected category's change is ranked among the changes of all
categories with at least 20 tracked applications over the same
calendar windows. For the revenue-share change we compare monthly new
listings in the twelve months before and after the effective date,
using snapshot creation dates for the full application universe so
that panel tracking plays no role. Finally, every series used in the
event studies is screened for temporary, later-recovered drops in detected installations (at least 5\% and 200 installations below the running maximum, counting only drops that later return to that maximum), the signature of a detection lapse (a genuine decline does not reverse), with episodes near window endpoints inspected individually.

\noindent
\textbf{\textit{RQ5: Survival and lifecycle.}}
For entry-observed applications, exit is defined as disappearance from
tracking more than eight weeks before the end of the panel, and
durations are ages in weeks. Applications still tracked when the
panel ends are treated as censored: we know they survived at least
that long but not how much longer. Detection can also lapse temporarily,
for example when an application changes its storefront code and
Store Leads has not yet attributed the new signature; an application
therefore counts as exited only if it stays undetected to the end of
the panel, so a reappearance cancels the exit. Because tracking exit
still conflates delisting with loss of storefront presence, we
validate it against snapshot status and repeat every cohort
comparison under a strict definition (exit confirmed Inactive in the
snapshot, or occurring after the snapshot date). We estimate Kaplan--Meier survival curves
by entry cohort and a Cox proportional hazards model with category
concentration at entry, freemium pricing, early review traction
measured strictly within the first 26 weeks of life, and entry year.

\noindent
\textbf{\textit{RQ6: Early-warning prediction.}}
Using entry-observed applications with at least 104 weeks of potential
follow-up, we predict exit within 104 weeks of creation from features
computed strictly within the first 26 weeks: installations reached,
installations added, whether any app-store review was received, the
rating, the pricing model, weeks tracked, and category size and
concentration at entry. We use logistic regression for
interpretability and report five-fold cross-validated AUC for both
exit definitions. AUC is the probability that the model ranks a
randomly chosen exiting application as riskier than a randomly chosen
surviving one, so 0.5 is chance and 1.0 is perfect; cross-validation
scores the model only on applications it was not fitted on.

\section{Results}\label{sec:results}

\subsection{RQ1: Dynamics and Scale of the Shopify App Ecosystem}\label{sec:rq1}

\paragraph{Ecosystem Scale}
The dataset contained 24,826 applications and 2,701,805 Shopify stores (live merchant stores only; development stores are not counted). Of the total applications, 16,698 (67.3\%) were marked as Active, indicating availability in the Shopify App Store at the time of data collection. The remaining 8,128 applications (32.7\%) had been removed or delisted. However, the majority of active applications showed no recorded installations. Only 4,213 active applications (17.0\% of the total) had recorded at least one installation, whilst 12,485 active applications (74.8\% of active apps) showed zero installs. This pattern likely reflects both genuine difficulty in achieving market entry and measurement limitations inherent in web scraping methodologies, which cannot detect installations when Store Leads fails to identify the application script in merchant storefronts.
Review activity provides a check on this split. Of the 12,485 zero-install active applications, 6,452 (51.7\%) have received at least one review in their lifetime, and 1,013 (8.1\%) received one in the 90 days before the snapshot; an application that is being reviewed is installed somewhere, so for at least this last group the zero reflects the detection limit. The large majority of zero-install applications, however, show no recent review activity, consistent with genuinely minimal adoption.

\begin{figure}[htbp]
\centering
\includegraphics[width=0.75\textwidth]{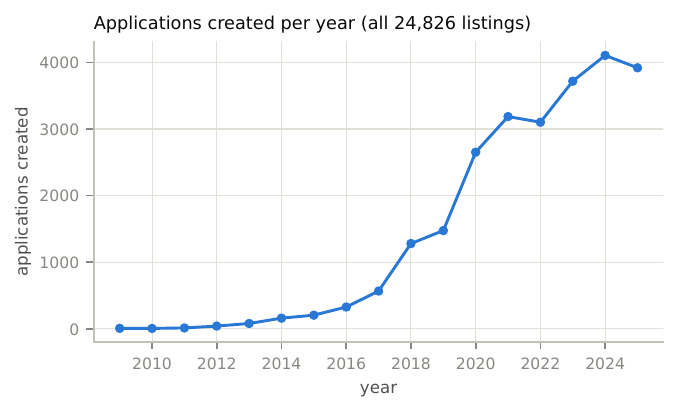}
\caption{Application creation timeline from 2009 to 2025 (all 24,826 listings, snapshot creation dates), showing more than twentyfold growth since 2015. The final point (2025) covers January to 28 September only; on that partial year the count (3,920) sits below the 2024 total (4,106) whilst the implied full-year pace is above it.}
\label{fig:app_timeline}
\end{figure}

Active applications were distributed across 231 primary functional categories. We identified 55 categories containing 100 or more applications and 126 categories containing 20 or more applications. The largest category, analytics, contained 655 applications, followed by shipping (591), discounts (561), chat (500), and dropshipping (447). The median category size was 27 applications, with the 25th percentile at 3 applications and the 75th percentile at 97 applications. 94 of the 231 categories (40.7\%) contained fewer than 10 applications.\\
\noindent
\textbf{Economic Significance}
Stores collectively generated an estimated \$706.3 billion in annual sales. These figures derive from traffic models rather than reported financials and represent merchant e-commerce activity, not app subscription revenue (which cannot be estimated due to unavailable pricing tier adoption data). Revenue distribution was highly skewed: the median store generated \$6,000 annually whilst the mean was \$261,435. At the 25th percentile, stores generated \$600 annually, whilst at the 99th percentile stores reached \$4.0 million. The maximum estimated revenue exceeded \$10 billion for a single domain; the extreme upper tail is dominated by merchandise stores attached to very high-traffic media and brand websites, where the traffic model inflates the estimate well beyond plausible storefront sales, so top-of-distribution values should be read as artefacts of the estimation method.\\
\noindent
\textbf{Application Adoption Patterns}
Stores varied substantially in application deployment. The median store installed 3 applications (mean: 4.0). A substantial portion (515,206, 19.1\%) installed zero applications. Most stores (1,457,466, 53.9\%) used 1 to 5 applications. Stores with 6 to 10 applications represented 18.0\% (486,776), whilst those with 11 to 20 applications accounted for 8.1\% (220,071). Only 22,286 stores (0.8\%) installed more than 20 applications (maximum: 141).
Install distribution across applications showed extreme concentration. The median install count was zero and the 75th percentile was 2 installs. Store Leads detects installations through storefront scripts, potentially undercounting private integrations, direct partnerships, or applications with limited client-side presence. Zero-install rates varied substantially by category, exceeding 95\% in technical infrastructure categories (payment providers, ERP, accounting) whilst remaining lower in consumer-facing categories, suggesting backend integrations are particularly difficult to detect.\\
\noindent
\textbf{Temporal Trends}
Application creation accelerated substantially over time, as shown in Figure~\ref{fig:app_timeline}. Among applications created since 2015, annual counts increased from 203 in 2015 to 4,106 in 2024, representing more than a twentyfold increase over nine years. The peak year for application creation was 2024, followed among complete years by 2023 (3,719 applications) and 2021 (3,187 applications). The year 2025, represented only partially in the dataset (data collected 28 September 2025), had already recorded 3,920 new applications, suggesting continued growth. Store creation followed a similar trajectory, with 7,622 new stores in 2015 rising to 466,906 in 2024.\footnote{The figure of 76,062 reported in the original study is a transcription error: the domains export gives 7,622 stores created in 2015, and neither the archived export nor any adjacent year contains a value near 76,062.}

The temporal span of the dataset extended 16.3 years for applications (earliest created 2 June 2009) and 19.3 years for stores (earliest created 2 June 2006). This longitudinal coverage enabled examination of first-mover dynamics in RQ3. The data confirmed that the Shopify ecosystem matured substantially after 2015, with both supply-side (application development) and demand-side (merchant adoption) growth accelerating in recent years.\\
\noindent
\textbf{Geographic Distribution}
Stores exhibited geographic concentration in a small number of markets. The United States accounted for 1,035,557 stores (38.3\% of the total), making it the dominant market by a substantial margin. The United Kingdom ranked second with 208,724 stores (7.7\%), followed by Canada (137,644 stores, 5.1\%), India (130,523 stores, 4.8\%), and Australia (126,226 stores, 4.7\%). The top five countries collectively represented 60.7\% of all stores in the dataset. Beyond these leading markets, adoption was distributed across more than 100 additional countries, though individual counts remained modest. This geographic pattern indicates that whilst Shopify operates as a global platform, the majority of merchant activity concentrates in English-speaking markets and India.\\
\noindent
\textbf{Pricing Models}
Among the 4,213 active applications with at least one installation, the majority employed freemium pricing strategies. A total of 2,922 applications (69.4\%) offered a minimum price of \$0, indicating free tiers or entirely free offerings. The remaining 1,291 applications (30.6\%) required payment from the outset, with minimum prices greater than zero. This distribution suggests that application developers predominantly rely on freemium models to reduce adoption barriers, likely converting users to paid tiers after initial trial periods or through premium feature upsells.\\
\noindent
\newline
\textbf{\textit{Summary}}
The Shopify app ecosystem comprises more than 16,000 active applications (of 24,826 ever listed) serving 2.7 million active merchant stores with an estimated \$706 billion in aggregate annual merchant revenue. Of 16,698 active applications, 74.8\% recorded zero installations at the time of data collection, though this figure likely reflects both measurement limitations in web scraping methodologies and genuine market entry barriers. Across all active applications, install counts remained heavily concentrated, with a median of zero and a 75th percentile of 2 installations; even among the 4,213 applications with at least one detected installation, the median was 123. A small number of applications dominated their categories whilst the majority showed limited recorded adoption. Merchants typically deployed small numbers of applications (median of 3), and geographic activity concentrated in the United States, which accounted for 38.3\% of stores. The ecosystem grew rapidly from 2015 to 2024, with application creation accelerating twentyfold. Most active applications employed freemium pricing models.

\subsection{RQ2: Concentration Within Categories}\label{sec:rq2}

\begin{figure}[ht]
  \centering
\includegraphics[width=0.8\textwidth]{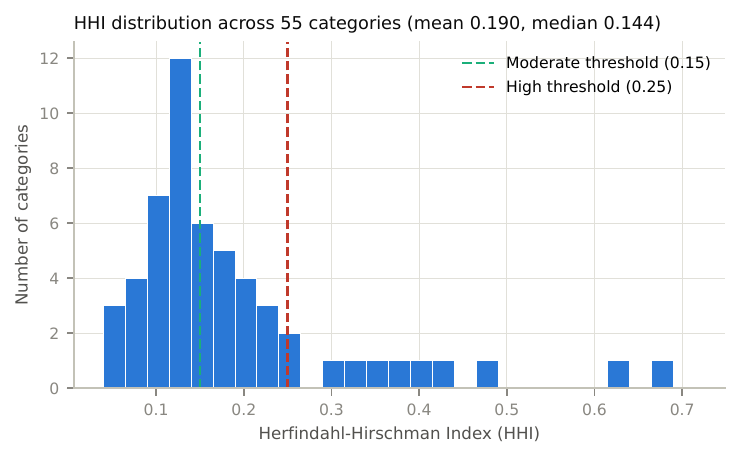}
  \caption{HHI distribution across 55 categories. Dashed lines show thresholds for moderate (0.15) and high (0.25) concentration.}
  \label{fig:hhi_dist}
\end{figure}

We analysed concentration patterns across the 55 categories that contained at least 20 active applications with market participation, that is, with at least one detected installation.
This selection differs from the category-size count reported in RQ1 (55 categories with at least 100 active applications, with or without installations): the equal count is a coincidence, and only 38 categories appear in both sets.
These categories collectively represented the competitive landscape of the Shopify app ecosystem, spanning functional areas from product reviews and email marketing to logistics and payments.\\
\noindent
\textbf{Overall Concentration Patterns}
Concentration varied substantially across categories, indicating that winner-take-all dynamics do not uniformly characterise the ecosystem. The mean Herfindahl-Hirschman Index across all analysed categories was 0.190, with a median of 0.144. Mean top-5 market share was 71.2\%, whilst the median was 71.1\%. Figure~\ref{fig:hhi_dist} presents the distribution of HHI values across all categories. The distribution exhibits right skew, with most categories clustering in the low to moderate range (HHI between 0.05 and 0.20) whilst a smaller number of categories extend into high concentration territory (HHI above 0.25). This pattern suggests that whilst leading applications typically capture the majority of adoption within their categories, a substantial portion of the market remains available to competitors in most functional areas.

Based on conventional HHI thresholds \cite{doj2010merger}, we identified 11 categories (20.0\%) exhibiting high concentration (HHI $\geq$ 0.25), 16 categories (29.1\%) exhibiting moderate concentration (0.15 $\leq$ HHI $<$ 0.25), and 28 categories (50.9\%) exhibiting low concentration (HHI $<$ 0.15).
Table~\ref{tab:catconc} lists every analysed category with its number of active applications, total detected installations, HHI, and the market shares of its leading applications, sorted from most to least concentrated, so that the competitive state of any category can be read directly.

\begin{table}[p]
  \centering
  \caption{Concentration by category (55 categories with at least 20
  active applications with installations, September 2025 snapshot).
  Top-1 and Top-5 are the installation shares of the leading and the
  five leading applications; Leading application is the application
  with the most installations; Level applies the HHI thresholds of
  0.15 and 0.25. The 11 High rows are the winner-take-all 20\%
  discussed in the text.}
  \label{tab:catconc}
  \footnotesize
  \setlength{\tabcolsep}{3pt}
  \begin{tabular}{lrrrrrll}
\toprule
Category & Apps & Installs & HHI & Top-1 (\%) & Top-5 (\%) & Leading application & Level \\
\midrule
social proof & 53 & 281,351 & 0.678 & 81.9 & 93.6 & Instafeed & High \\
design elements - other & 42 & 19,590 & 0.634 & 79.4 & 89.6 & Fontify & High \\
chat & 145 & 560,143 & 0.467 & 67.8 & 82.4 & Shopify Inbox & High \\
wishlists & 44 & 80,322 & 0.428 & 64.4 & 84.0 & Swym Wishlist Plus & High \\
search and filters & 36 & 41,288 & 0.393 & 60.9 & 87.0 & Boost AI Search \& Filter & High \\
pricing optimization & 24 & 8,416 & 0.382 & 58.8 & 91.5 & Intelligems & High \\
images and media - other & 22 & 4,796 & 0.360 & 54.9 & 91.7 & Magic Zoom Plus & High \\
product comparison & 54 & 82,412 & 0.332 & 54.2 & 88.7 & Kiwi Size Chart \& Recom\dots & High \\
custom file upload & 23 & 13,058 & 0.313 & 50.2 & 94.2 & Uploadly & High \\
anti theft & 29 & 19,541 & 0.263 & 45.6 & 80.8 & Cozy AntiTheft & High \\
product reviews & 131 & 1,015,412 & 0.253 & 47.7 & 72.9 & Judge.me Product Review\dots & High \\
forms & 33 & 95,171 & 0.231 & 37.6 & 85.1 & Powerful Contact Form B\dots & Moderate \\
blogs & 22 & 8,585 & 0.219 & 35.6 & 89.9 & AfterShip Page Builder & Moderate \\
order tracking & 37 & 245,935 & 0.219 & 34.1 & 87.7 & 17TRACK Order Tracking & Moderate \\
banners & 111 & 465,676 & 0.212 & 42.5 & 74.4 & Klarna On & Moderate \\
page builder & 71 & 495,419 & 0.210 & 41.0 & 78.0 & PageFly Landing Page Bu\dots & Moderate \\
subscriptions & 40 & 138,057 & 0.202 & 34.2 & 81.6 & Recharge Subscriptions\dots & Moderate \\
email marketing & 70 & 1,033,276 & 0.191 & 36.3 & 80.8 & Klaviyo & Moderate \\
store locator & 31 & 32,065 & 0.188 & 38.5 & 72.5 & Stockist Store Locator & Moderate \\
stock alerts & 54 & 91,972 & 0.172 & 27.4 & 77.6 & AMP Back in Stock & Moderate \\
countdown timer & 43 & 115,920 & 0.170 & 26.3 & 77.9 & Essential Countdown Tim\dots & Moderate \\
currency and translation & 48 & 379,155 & 0.169 & 34.4 & 70.9 & BUCKS Currency Converte\dots & Moderate \\
wholesale & 33 & 44,407 & 0.166 & 34.7 & 70.7 & Faire & Moderate \\
abandoned cart & 72 & 143,557 & 0.156 & 30.1 & 73.1 & Dondy & Moderate \\
affiliate programs & 64 & 245,526 & 0.155 & 24.0 & 77.0 & BixGrow Affiliate Marke\dots & Moderate \\
faq & 28 & 23,720 & 0.152 & 30.2 & 74.4 & SB & Moderate \\
pre-orders & 41 & 145,575 & 0.151 & 29.9 & 75.4 & PreOrder Globo & Moderate \\
dropshipping & 177 & 528,074 & 0.144 & 27.0 & 74.5 & Printful & Low \\
animation and effects & 30 & 5,132 & 0.142 & 28.0 & 71.1 & Easy Background Music & Low \\
legal & 48 & 35,830 & 0.135 & 31.0 & 63.8 & Unicorn Age Gate & Low \\
delivery and pickup & 63 & 59,059 & 0.135 & 27.5 & 67.7 & Bird Pickup Delivery Da\dots & Low \\
badges and icons & 61 & 151,258 & 0.133 & 28.3 & 69.0 & Conversion Bear Trust B\dots & Low \\
product display - other & 46 & 15,169 & 0.130 & 26.0 & 69.8 & FOLDER & Low \\
print on demand (pod) & 93 & 93,468 & 0.127 & 23.3 & 65.1 & Customily Product Perso\dots & Low \\
shipping rates & 27 & 12,002 & 0.127 & 22.3 & 74.9 & Shipping \& Delivery & Low \\
loyalty and rewards & 85 & 231,019 & 0.126 & 25.7 & 68.0 & Smile & Low \\
cart customization & 159 & 286,274 & 0.123 & 30.6 & 56.7 & AMP & Low \\
analytics & 100 & 153,062 & 0.119 & 25.6 & 57.4 & Lucky Orange Heatmaps \&\dots & Low \\
shipping & 27 & 21,761 & 0.119 & 18.9 & 71.0 & Colissimo Official & Low \\
navigation and menus & 46 & 88,986 & 0.118 & 22.8 & 70.6 & Smart Product Filter \&\dots & Low \\
accounts and login & 64 & 65,348 & 0.117 & 27.5 & 59.9 & Locksmith & Low \\
donations & 30 & 8,691 & 0.113 & 22.6 & 64.3 & GoodAPI Plant Tree Clea\dots & Low \\
seo & 60 & 263,468 & 0.111 & 20.8 & 66.6 & Avada AI SEO Image Opti\dots & Low \\
pricing quotes & 56 & 36,020 & 0.110 & 26.0 & 60.0 & Q & Low \\
returns and exchanges & 33 & 31,739 & 0.104 & 17.4 & 66.3 & AfterShip Returns \& Exc\dots & Low \\
video and livestream & 60 & 42,288 & 0.095 & 21.7 & 57.2 & Tolstoy AI Shoppable Vi\dots & Low \\
product bundles & 86 & 236,599 & 0.093 & 16.7 & 63.3 & Bundler & Low \\
3d/ar/vr & 30 & 1,360 & 0.093 & 18.4 & 59.0 & Picture It & Low \\
pop-ups & 93 & 241,629 & 0.089 & 20.4 & 52.9 & Trustoo Pop ups, Email\dots & Low \\
event booking & 35 & 62,466 & 0.081 & 12.9 & 52.8 & Booking App Schedule Co\dots & Low \\
product variants & 114 & 456,099 & 0.080 & 16.0 & 53.5 & King Product Options \&\dots & Low \\
image gallery & 65 & 73,572 & 0.076 & 15.3 & 51.4 & Socialwidget & Low \\
upsell and cross-sell & 149 & 237,012 & 0.059 & 16.1 & 43.4 & Selleasy & Low \\
ads & 74 & 103,569 & 0.052 & 11.7 & 40.3 & AdRoll Marketing \& Adve\dots & Low \\
discounts & 184 & 166,972 & 0.047 & 11.4 & 42.9 & Discount Ninja Promo En\dots & Low \\
\bottomrule
\end{tabular}

\end{table}
The majority of categories therefore operated in competitive or moderately competitive regimes rather than monopolistic structures. This finding indicates that whilst dominant applications exist in many categories, the ecosystem maintains sufficient diversity that new entrants face varied competitive conditions depending on their chosen functional area.\\
\noindent
\textbf{Highly Concentrated Categories}
The 11 highly concentrated categories (the High rows of Table~\ref{tab:catconc}) exhibited strong winner-take-all dynamics. The most concentrated category, social proof, recorded an HHI of 0.678, with the leading application capturing 81.9\% market share and the top five applications collectively holding 93.6\%. Other highly concentrated categories included design elements (HHI = 0.634), chat (HHI = 0.467), wishlists (HHI = 0.428), and search and filters (HHI = 0.393). These categories exhibited clear market dominance, with single applications controlling majority positions.

Among the top 20 categories ranked by total installation volume, three exceeded the high concentration threshold. Social proof exhibited the highest concentration (HHI = 0.678), with Instafeed dominating with 81.9\% market share. Chat also qualified as highly concentrated (HHI = 0.467), with Shopify Inbox holding 67.8\% market share. Product reviews (HHI = 0.253) also exceeded the threshold, with Judge.me Product Reviews App capturing 47.7\% of installations. The remaining 17 categories among the top 20 operated in low or moderate concentration regimes, indicating that the largest and most economically significant functional areas generally maintained competitive structures.\\
\noindent
\textbf{Competitive Categories}
At the opposite end of the distribution, the least concentrated categories exhibited fragmented market structures with no dominant providers. The category with the lowest concentration, discounts, recorded an HHI of 0.047, with the leading application (Discount Ninja) capturing 11.4\% market share and the top five collectively holding 42.9\%. Other highly competitive categories included ads (HHI = 0.052), upsell and cross-sell (HHI = 0.059), image gallery (HHI = 0.076), and product variants (HHI = 0.080). These categories featured dozens or hundreds of applications competing for adoption, with no single provider establishing clear dominance.

Among the top 20 categories by installation volume, several major functional areas remained highly competitive despite their economic importance. The upsell and cross-sell category, with 149 active applications and 237,012 total installations, exhibited an HHI of only 0.059, with the leading application (Selleasy) capturing 16.1\% market share. Similarly, the discounts category (HHI = 0.047) maintained highly competitive structures despite substantial total adoption. Email marketing (HHI = 0.191) and page builder (HHI = 0.210) operated in moderate concentration regimes, indicating viable competition among multiple providers.

\paragraph{Concentration and Category Characteristics}
We examined whether concentration varied systematically with category size, measured by the number of active applications.
The mean HHI decreased with category size. Smaller categories exhibited higher concentration, whilst larger categories sustained more competitors at viable market shares. This inverse relationship between category size and concentration contradicts the expectation that larger markets would exhibit stronger winner-take-all effects due to network externalities or economies of scale. Instead, larger categories sustained more competitors at viable market shares, suggesting that functional diversity, merchant heterogeneity, or niche differentiation opportunities expand more rapidly than concentration forces in growing categories.

We also examined whether concentration correlated with total category adoption, measured by aggregate installations across all applications in a category. No clear systematic relationship emerged. The 30 categories with fewer than 100,000 total installations spanned HHI values from 0.08 to 0.63, whilst the two categories with more than 1 million installations recorded 0.19 and 0.25. High adoption categories included both highly concentrated examples (social proof with 281,351 installations and HHI = 0.678) and competitive examples (upsell and cross-sell with 237,012 installations and HHI = 0.059). This pattern indicates that market size alone does not determine competitive structure.

Concentration is also not a residue of early-mover advantage. Across
the categories that enter both analyses, HHI is uncorrelated with the
entry-timing correlations of RQ3 (Spearman $\rho = 0.01$ for
velocity and $-0.05$ for rate, recomputed as in
Section~\ref{sec:rq123}), and all ten highly concentrated categories
in that comparison favour later entrants on both growth measures.
Dominant positions in concentrated categories therefore coexist with
the late-mover pattern: a category
concentrates around one strong application whilst later entrants
still outgrow earlier ones across the rest of the field.\\
\noindent
\textbf{Market Share Distribution Within Categories}
Examining the distribution of market shares within individual categories revealed additional structure beyond aggregate HHI values. In the highly concentrated categories, the leading application captured between 46\% and 82\% market share, and the top five applications collectively held 73\% to 94\%. In moderately concentrated categories, the leader held 24\% to 43\% and the top five 71\% to 90\%. In low concentration categories, the leader held 11\% to 31\% and the top five 40\% to 75\%, so the leader's share grades continuously across the bands whilst never approaching the dominance seen at the top of the distribution.

The relationship between top-1 market share and top-5 market share varied across categories. In winner-take-all categories, a single dominant application often held market share approaching that of the entire top five combined, indicating weak secondary competition. In more fragmented categories, the top five applications held substantially more combined share than the leader alone, indicating that multiple competitors operated at comparable scale. This pattern suggests that concentration manifests through different mechanisms: some categories exhibit single-firm dominance, whilst others feature oligopolistic structures with several large competitors and a fragmented tail.\\
\noindent
\newline
\textbf{\textit{Summary.}}
Concentration within Shopify app categories varies substantially depending on functional area. Whilst 20\% of categories exhibit winner-take-all dynamics with dominant providers capturing majority market share, more than half of all categories operate in the low-concentration regime (HHI below 0.15), where no leading application reaches a third of the market. Larger categories, measured by number of competing applications, systematically exhibit lower concentration, suggesting that market growth enables rather than forecloses competition. The most economically significant categories span the full range of concentration levels, with some (social proof, chat, product reviews) dominated by incumbents and others (upsell and cross-sell, discounts, email marketing) maintaining diverse competitive fields. These patterns indicate that opportunities for new applications depend critically on category selection, with competitive functional areas offering substantially greater prospects for market entry than monopolised categories.

\subsection{RQ3: Early Mover Growth Dynamics}\label{sec:rq3}

With RQ3 we aimed at understanding whether apps entering categories earlier maintain growth advantages over time. Whilst cumulative install counts mechanically favour older applications through longer market exposure, recent growth patterns reveal whether early movers sustain competitive momentum or face erosion from newer entrants.
We analysed 3,456 applications with at least 180 days of market tenure, a restriction that ensures growth patterns had stabilised beyond initial launch volatility; the category-level results below rest on the 2,809 of these applications that fall within the 50 categories containing at least 20 qualifying applications.

We decomposed growth into two dimensions: velocity (absolute growth measured by installations in the last 90 days) and rate (proportional growth calculated as recent installations divided by total installations). This separation distinguishes visibility advantages from momentum effects. We calculated Spearman rank correlations between app creation timing and both growth dimensions, applying FDR correction (Benjamini-Hochberg procedure, $q < 0.05$) to control for multiple comparisons across categories.\\
\noindent
\textit{Growth Patterns Favour Late Entrants.} Contrary to expectations of persistent first mover advantages, late entrants exhibited stronger growth across both dimensions in the majority of categories. Of 50 categories, 45 (90.0\%) showed positive velocity correlations, indicating that later-created apps achieved higher absolute growth than earlier-created apps. Similarly, 48 categories (96.0\%) exhibited positive rate correlations, indicating faster proportional growth amongst late entrants. Nine velocity correlations and eight rate correlations reached statistical significance after FDR correction.

\begin{figure}[t]
\centering
\includegraphics[width=0.75\textwidth]{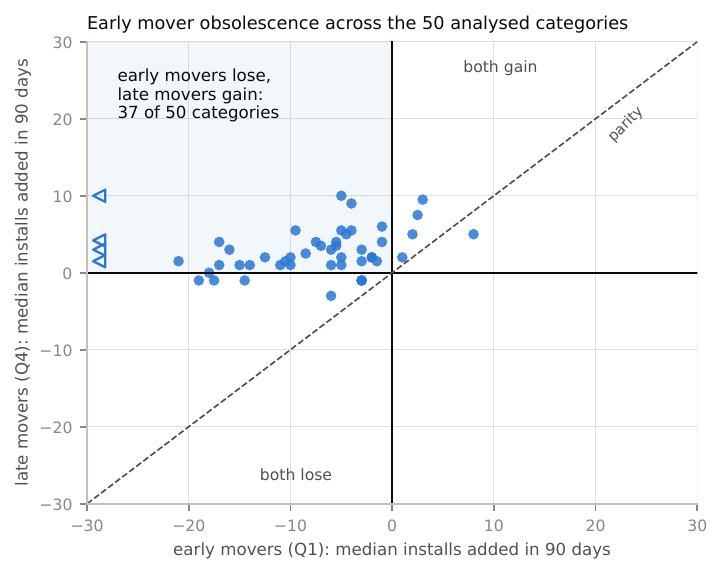}
\caption{Early mover obsolescence across the 50 analysed categories.
Each point is a category, comparing the median net installations
added over 90 days by its earliest quartile of applications
(horizontal) with the same quantity for its latest quartile
(vertical); negative values mean the median application lost
installations. Points above the dashed parity line are categories
whose median late mover outgrows its median early mover. In 37 of 50
categories (74\%, shaded region) the median early mover loses
installations whilst the median late mover gains them. Four
categories with early-mover medians beyond the axis range ($-31.5$
to $-87$; order tracking the most extreme) are shown as
left-pointing markers at the edge, at their late-mover values.}
\label{fig:rq3_obsolescence}
\end{figure}

The pattern proved consistent across major functional areas. Email marketing, the largest category by installation volume, showed velocity correlation of $\rho = 0.261$ and rate correlation of $\rho = 0.203$. Product reviews exhibited $\rho = 0.186$ for velocity and $\rho = 0.066$ for rate. Chat applications showed $\rho = 0.203$ for both dimensions. Only five categories showed negative velocity correlations, indicating early mover growth advantages, and only two showed negative rate correlations.\\
\noindent
\textbf{Early Mover Obsolescence, Not Late Mover Advantage.} Examination of median growth values revealed that the observed pattern reflects early mover decline rather than exceptional late mover performance. Figure~\ref{fig:rq3_obsolescence} presents median velocity for early movers (first quartile by creation date) versus late movers (fourth quartile) across all 50 categories. The concentration of 37 categories (74\%) in the upper-left quadrant demonstrates systematic early mover obsolescence: early movers on net lost detected installations (negative velocity) whilst late movers gained them (positive velocity). Of 50 categories, 45 exhibited negative median velocity for early movers, a net decline in detected installations, whilst 42 exhibited positive median velocity for late movers.

In product reviews, early movers recorded median velocity of $-12.5$ installations per 90 days whilst late movers recorded $+2$ installations. Email marketing showed $-1.5$ for early movers and $+1.5$ for late movers. Order tracking exhibited the starkest contrast, with early movers at $-87$ and late movers at $+1.5$. This pattern indicates lifecycle dynamics in which early movers on net lose detected installations whilst late movers gain them. Store Leads data track net changes in visible installations, detecting both additions and removals as merchants modify their storefronts. Negative growth values therefore record net losses of detected installations; Section~\ref{sec:threats} discusses the residual risk that an individual decline reflects a detection lapse.

We classified categories into competitive dynamics based on correlation patterns. Only one category (2.0\%) exhibited strong first mover advantage (negative correlations for both velocity and rate). Four categories (8.0\%) showed the eroding first mover advantage pattern (negative velocity correlation, positive rate correlation), where early movers maintained absolute growth advantages despite late movers achieving faster proportional growth. The remaining 44 categories (88.0\%) favoured late entrants on both dimensions, whilst one category showed no clear pattern. This classification rests on the correlations across all applications in a category and not on the quartile medians of Figure~\ref{fig:rq3_obsolescence}, so a category can favour late entrants overall whilst its quartile medians sit near parity.

Whilst the systematic pattern favours late entrants, individual outcomes vary within categories. Product reviews illustrates this heterogeneity: the category exhibits high concentration with a dominant early mover sustaining strong growth, whilst other early entrants experience typical obsolescence patterns. This demonstrates that early entry combined with sustained quality and continuous innovation can maintain competitive positions, though the category-level pattern of late mover advantage persists.

These exceptions are frequent enough to characterise. Pooling the
earliest quartile of entrants across the recomputed categories gives
849 early movers, of which 254 (30\%) were still adding installations
on net in the 90 days before the snapshot. What separates this
growing minority from the declining majority is current standing, not
vintage: resilient and declining early movers have the same median
age (8.1 years), but the resilient ones hold larger installed bases
(median 1,602 versus 466 installations), carry more reviews (110
versus 38) at higher ratings (4.80 versus 4.70), more often offer a
free tier (67\% versus 55\%), and are nearly three times as likely to
rank among their category's five largest applications (29\% versus
10\%). Market position carries much of this: 54\% of early movers in
a category's top five are still growing, against 25\% of the rest,
and leadership itself is often held by an early mover: the largest
application is an early mover in 31 of the 54 categories, and in 21
of those 31 it is still growing. The leadership picture therefore
splits roughly forty--twenty--forty: in 39\% of categories the leader
is an early mover that is still growing, in 19\% an early mover that
has stopped growing, and in 43\% an application from outside the
earliest quartile.
The quality signals matter beyond position, however, because among
early movers outside the top five the growing minority still shows
twice the review base (60 versus 30) at higher ratings. Judge.me is
the clearest case: created in 2015 and in the earliest quartile of
product reviews, it grew from roughly 11,000 detected installations
in February 2019 to 531,000 in March 2026, in a category whose
correlations nonetheless favour later entrants. Nor is product
reviews unusual in hosting such an exception: in 44 of the 54
categories (of the 55 concentration categories, one falls below 20
applications once the 180-day maturity filter is applied), at least
one early mover among the five largest
applications is still adding installations, so a resilient incumbent
at the top is the norm even where the category-level pattern favours
later entrants. Early entry itself
protects nothing; early entrants that converted their head start into
leadership, review volume, and a free tier continue to grow, and
obsolescence concentrates among those that did not. Why this minority
resists obsolescence, whether through product quality, support,
pricing, or organisational factors, cannot be answered from public
listing data and is left to future work.\\
\noindent
\newline
\textbf{No Evidence of Momentum Shifts.}
To test whether late movers exhibited accelerating growth that might eventually reverse early mover advantages, we compared recent momentum (30-day installations) against longer-term averages (90-day installations divided by three). Apps with 30-day growth exceeding their 90-day average were classified as accelerating. Within each category, we compared acceleration rates between early movers and late movers using Mann-Whitney U tests with FDR correction.

Of 50 categories, 21 (42.0\%) showed higher acceleration amongst late movers, whilst 29 showed higher acceleration amongst early movers. However, no category reached statistical significance after FDR correction. Mean Cohen's $d$ was $-0.080$, indicating a negligible effect size. The proportion of apps exhibiting acceleration was 36.7\% for early movers and 38.3\% for late movers, a difference of 1.6 percentage points. These results provide no evidence of directional momentum shifts that would alter competitive dynamics over time.\\
\noindent
\textit{\textbf{Illustrative Catch-Up Scenarios.}}
To contextualise the magnitude of growth differences, we projected forward under three mathematical scenarios, emphasising that these represent illustrative bounds rather than predictions. We calculated baseline installation counts and growth metrics for early and late movers in each category, then projected forward over 20 quarters (five years) under exponential growth (constant rates), linear growth (constant absolute additions), and decay (rates declining linearly to zero). These scenarios assume growth patterns persist without market saturation, competitive responses, quality changes, or platform algorithm shifts.

Under the realistic decay scenario, which assumes growth rates gradually decline to zero over the projection horizon, only five categories provided sufficient positive growth data for projection. Among these, one category (20.0\%) showed late movers reaching 50\% of early mover size within two years, whilst four (80.0\%) showed late movers failing to reach this threshold within five years. Mean ratio after 24 months was 0.58, indicating late movers reached 57.8\% of early mover installation volumes under these assumptions. Median time to parity was indeterminate due to insufficient categories achieving the threshold.

However, the limited applicability of these projections (five categories rather than 50) reflects the prevalence of negative growth amongst early movers, which violates the mathematical assumptions required for forward projection. The projections therefore provide limited insight beyond confirming that observed growth differentials, if sustained, would require extended periods for convergence.\\
\noindent
\newline
\textbf{\textit{Summary.}} Recent growth patterns systematically favour late entrants over early movers across 88\% of analysed categories. Stated plainly, in most categories the median early mover now loses detected installations over a 90-day window whilst the median late mover gains them, despite the early movers' years of accumulated visibility. This pattern reflects early mover obsolescence, with established applications on net losing detected installations whilst newer alternatives capture growth. The dynamics suggest app markets remain contestable despite installed base advantages, as detected adoption shifts towards newer solutions over time. No evidence emerged of accelerating late mover momentum that would further alter competitive dynamics. These findings indicate that entry timing confers advantages through recency rather than incumbency, contradicting traditional first mover advantage theory in platform app ecosystems.

\subsection{Replication Bridge and Longitudinal Context}\label{sec:rq123}

We re-derive the preceding results from the raw snapshot with
independent code. The scale funnel is reproduced exactly (24,826
applications, 16,698 active, 4,213 active with installations, 231
primary categories over active applications), as is every published
concentration statistic: the 55 analysed categories, mean HHI 0.190,
median 0.144, mean top-five share 71.2\%, and the classification of 11
categories as highly, 16 as moderately, and 28 as lowly concentrated.
For the entry-timing analysis, the stated inclusion criteria yield 4,005 mature applications, 3,314 of them inside 54 qualifying categories, against the original's 3,456, of which 2,809 fall inside its 50 categories. The difference traces to an additional, unstated filter in the original analysis: beyond the criteria above, the original computation required a computable acceleration ratio, which excluded mature applications with no installations in the trailing 90 days. Reapplying this filter reproduces the published sample exactly. Our recomputation retains
these zero-growth applications. The substantive result is insensitive
to the choice and slightly stronger in our recomputation: 94.4\% of
categories show a positive velocity correlation and 98.1\% a positive
rate correlation, against the published 90\% and 96\%. Finally, panel
and snapshot installation counts agree exactly at the overlapping
week, so the longitudinal analyses below rest on the same measurement
process as the original study.

\begin{figure}[t]
  \centering
  \includegraphics[width=\textwidth]{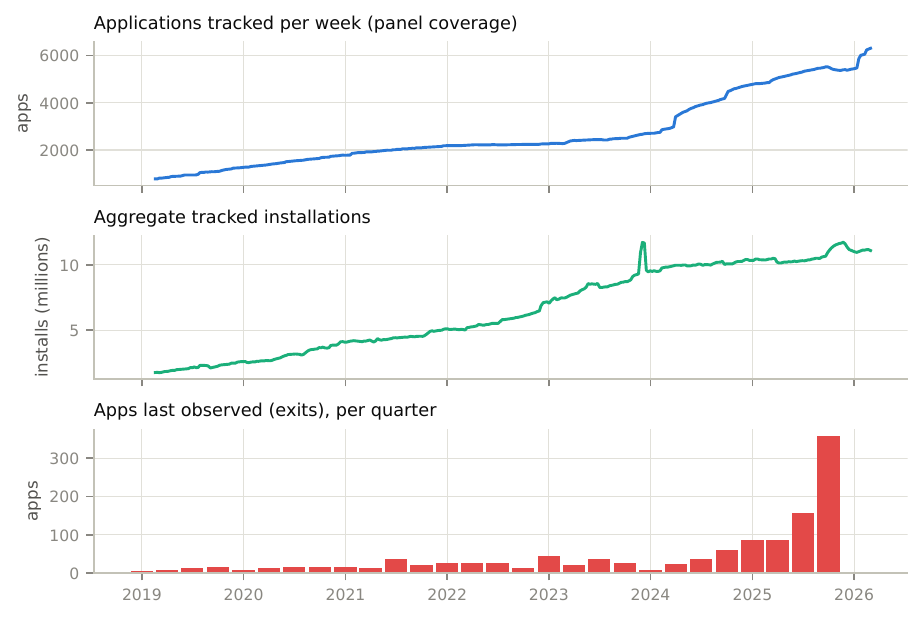}
  \caption{Panel backdrop, 2019--2026. Top: applications tracked per
  week (tracking coverage). Middle: aggregate tracked installations;
  the brief spike in late 2023 is a documented scraping glitch on a single application outside every event-study category, so no event window's series includes it. Bottom: applications last observed (exits from tracking)
  per quarter. The final two quarters are
  provisional, because fewer than 26 weeks of panel remain to
  confirm each absence: requiring 26 weeks of observed absence
  lowers the 2025Q3 count from 156 to 63 and leaves the 2025Q4
  count unconfirmed.}
  \label{fig:backdrop}
\end{figure}

\begin{figure}[t]
  \centering
  \includegraphics[width=\textwidth]{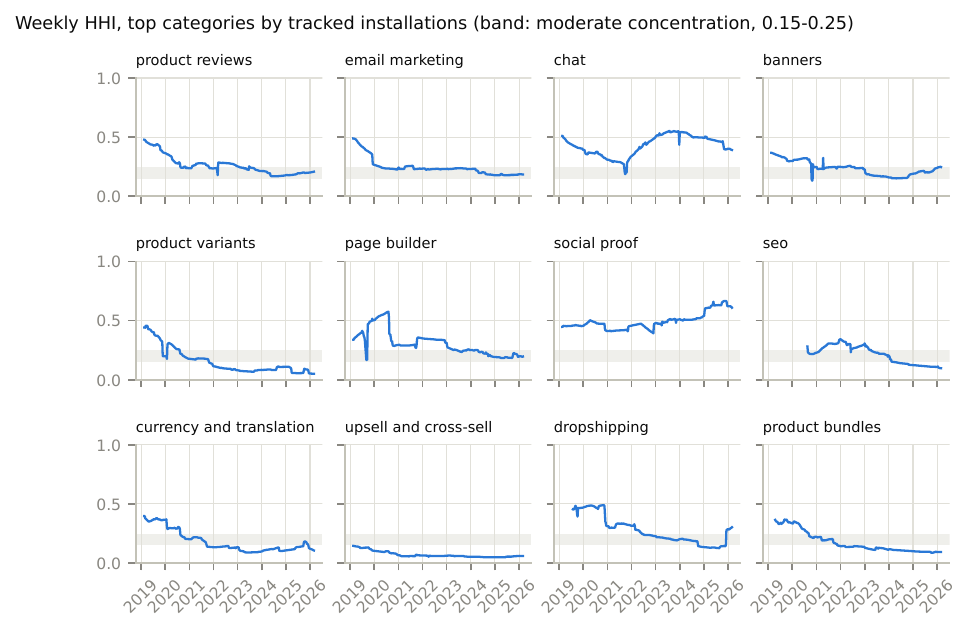}
  \caption{Weekly HHI for the twelve largest categories by tracked
  installations, 2019--2026, using the original study's category definitions. The shaded band marks moderate concentration (0.15--0.25). Each panel begins in the first week its category has at least ten tracked applications.}
  \label{fig:evolution}
\end{figure}

Over the panel period the ecosystem grew from 1.8M to 11.1M tracked
installations (Figure~\ref{fig:backdrop}). Concentration trajectories for the 68 categories with
sufficient tracked history show that 53 de-concentrated significantly over 2019--2026, 13 concentrated, and two showed no significant trend (Figure~\ref{fig:evolution}); the concentrating minority includes
chat, whose platform entry is examined next, and social proof,
whilst product reviews de-concentrated over the period as a whole
and re-concentrated only after the platform's exit
(Section~\ref{sec:rq4}). Market leadership itself
proved unstable: in 92\% of the 85 categories with at least 20 tracked
applications, the top application by installations changed at least
once over the seven years (median three changes; requiring the new
leader to persist for at least eight weeks), with discounts changing
leaders ten times. Contestability in this ecosystem extends to the top
of the ranking as well as to its tail.

\begin{figure}[t]
  \centering
  \includegraphics[width=\textwidth]{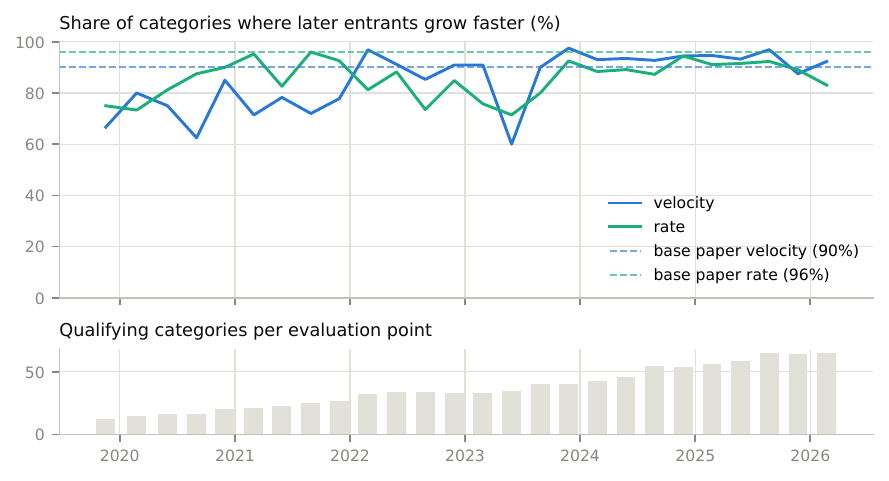}
  \caption{Share of categories in which later entrants grow faster,
  recomputed quarterly with the original inclusion rules (top), and
  the number of qualifying categories at each evaluation point
  (bottom). The dashed lines mark the original study's separate
  estimates: 90\% of categories with a positive velocity correlation
  and 96\% with a positive rate correlation.}
  \label{fig:rolling}
\end{figure}

The entry-timing result itself is not specific to the original
measurement window. Recomputing the entry-timing correlations on a
quarterly grid across the panel shows that the share of categories
favouring later entrants has a median of 90.5\% for velocity and
87.9\% for rate across 26 evaluation points spanning 2019--2026,
never falling below 60\%, and from 2024 onwards fluctuates around the
published estimates of 90\% and 96\% (velocity between 87\% and
97\%, rate between 83\% and 94\%) (Figure~\ref{fig:rolling}). The
original September 2025 estimates are typical of the period from
2022 onwards; the sparser early points, drawn from 12 to 27
categories per quarter, sit lower, with velocity shares between
62\% and 85\% before 2022.

The stability is also not an artefact of how growth is measured. The
original 90-day metrics exist because the snapshot only reports
installations over fixed trailing windows; with the weekly panel we
can vary the measure. Repeating the grid with a 26-week window, and
again with velocity measured as the fitted weekly trend over every
observation in the window instead of the difference between its two
endpoints, leaves the picture unchanged: the median share of
categories favouring later entrants is 90.7\% and 88.1\% for velocity
(against 90.5\% in the baseline) and 85.6\% and 86.0\% for rate
(against 87.9\%). The panel also allows a cleaner version of the
original momentum test, which had to infer acceleration from two
overlapping windows observed on a single date: measuring acceleration
directly, as an application's velocity over the most recent 13 weeks
minus its velocity over the preceding 13 weeks, and repeating the
early-versus-late comparison at every quarterly point, confirms the
published null result. Only 3 of 913 category-level tests are
significant after false-discovery-rate correction, 24 of the 26
evaluation points have none, and the mean effect size is negligible
(Cohen's $d = 0.08$, against $-0.08$ in the original; the sign difference is a convention artefact: the original computed $d$ as late movers minus early movers on a continuous acceleration ratio, whilst ours is early minus late on the accelerating share, so under a common convention both values are $+0.08$, agreeing in direction and negligible in size). Finally, the
detection lapses discussed in Section~\ref{sec:threats} do not drive
the result either: screening all 7,708 tracked series for the
drop-and-rebound signature of a lapse flags 10\% of applications, and
excluding every application whose growth window overlaps such an
episode leaves the medians essentially unchanged (88.5\% for velocity
and 87.0\% for rate), whilst the weakest evaluation point, a mid-panel dip in May 2023, improves from 60\% to 66\%, indicating that part of that dip came from measurement.

\subsection{RQ4: Platform Governance Events}\label{sec:rq4}

\begin{figure}[t]
  \centering
  \includegraphics[width=\textwidth]{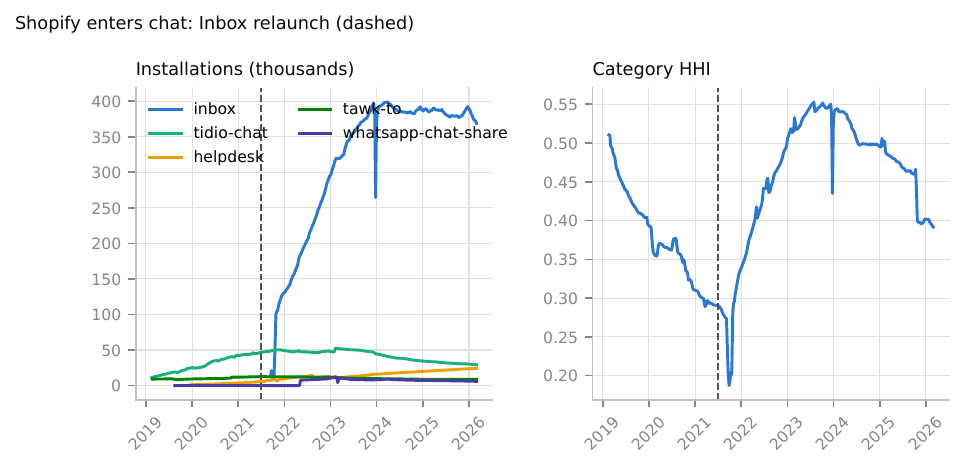}
  \caption{Shopify's entry into chat. Left: installations of Inbox and
  leading incumbents. Right: weekly category HHI. The dashed line
  marks the Inbox relaunch (July 2021).}
  \label{fig:inbox}
\end{figure}

\paragraph{Entry: Shopify Inbox.}
Before the relaunch of Shopify Inbox in July 2021 (a consolidation of
the earlier Shopify Chat and Ping products), chat had been
de-concentrating since tracking began in February 2019 (HHI declining
from 0.51 to 0.29).
Inbox reversed the trend: its installed base peaked just under
400,000 installations in March 2024, within three years of the
relaunch, and category HHI rose to a peak of 0.55
(Figure~\ref{fig:inbox}). Shopify Inbox's growth made chat one of the most concentrated categories in the original snapshot. Incumbent growth also slowed: Tidio, the
leading third-party incumbent, fell from a median of +193
installations per week in the 26 weeks before the relaunch to +48 in
the 26 weeks after. The permutation check tempers a causal reading of
that slowdown: chat's change in median incumbent growth ranks at the
32nd percentile of all category changes over the same calendar
windows, a slowdown, but within the range of contemporaneous
market-wide dynamics. The concentration reversal, which is driven by
Inbox's own share, is the robust component of this event.

\begin{figure}[t]
  \centering
  \includegraphics[width=\textwidth]{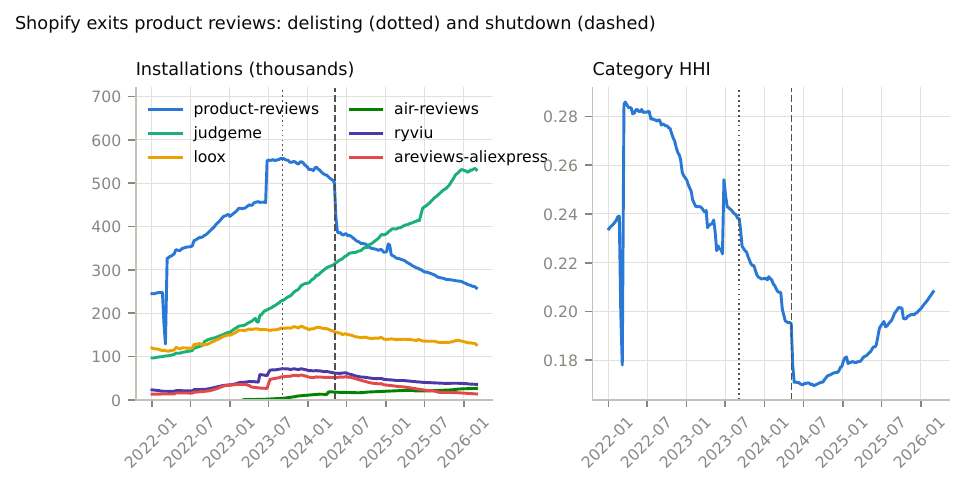}
  \caption{Shopify's exit from product reviews. Left: installations of
  the platform application and leading competitors. Right: weekly
  category HHI. Dotted line: delisting (September 2023); dashed line:
  shutdown (May 2024).}
  \label{fig:deprecation}
\end{figure}

\paragraph{Exit: the Product Reviews deprecation.}
The deprecation of Shopify's own Product Reviews application is, to
our knowledge, the first measured case of a platform owner exiting a
complementor category, and its central lesson is that only a minority of the released installed base reappeared as competitor adoption. After the
shutdown on 6 May 2024, the platform application lost 191,290 detected
installations within 52 weeks. Competitors gained 141,877
installations over the same window, but gross gains conflate the event
with pre-existing trends: the category leader Judge.me, whose gross
gains dominate, was already adding 133,637 installations in the 52
weeks before the shutdown and added fewer (94,530) in the 52 weeks
after it. That slowdown is confined to the post-shutdown year. The
leader's installed base grew in every calendar year of the panel; its
yearly additions rose in every year except 2024, when they were
essentially flat (112,704 against 113,492 in 2023), and 2025, outside
the event window, was its largest year (148,675 added). Netting each established competitor's own pre-event trend
leaves 32,851 event-attributable installations; adding every
installation gained by applications newly listed during the window
raises the ceiling to 60,684. Between 17\% and 32\% of the platform
application's lost installed base therefore reappeared as competitor
adoption within a year; the majority of the lost detected installed base did not reappear in any competitor within the year, consistent with much of the free default application's installed base having been inactive.
Category HHI reached its minimum fourteen weeks after the shutdown
and rose steadily thereafter as the platform application's share
dissolved (Figure~\ref{fig:deprecation}). Seen as market penetration,
the exit marks a turning point for the category as a whole: product
reviews' share of all tracked installations peaked at 15.4\%
(14.4\% excluding the single-week step described at the end of this
paragraph) two weeks after the delisting and fell to 12.1\% by the end of the panel. The
decline is fully accounted for by the platform application's
dissolving base; excluding it, third-party penetration rose from
8.7\% to 9.7\% over the same period. Within the third-party segment,
adoption concentrated on the leader throughout: Judge.me's share of
tracked category installations grew from 5.5\% in February 2019 to
17.6\% at the delisting and 39.7\% by the end of the panel. The
timing of the peak coincides
with the exit, and we make no causal claim. These numbers are not an
artefact of detection lapses: the screen described in
Section~\ref{sec:method} finds no recovered drop positioned to bias
the window endpoints, the platform application's decline never
reverses, Judge.me's series contains no recovered drawdown at all,
and the one episode straddling an endpoint (a competitor dip in
spring 2023, across the start of the pre-trend window) inflates that
competitor's pre-trend and therefore understates the recovery share,
making the 17\% lower bound conservative. The screen
is one-sided by construction: it detects drops that later recover, so
a permanent one-week upward step evades it. The platform
application's series contains one such step, +96,969 installations
(21.3\%) in the week of 25 June 2023, inside the pre-event window.
The event quantities are differences between window endpoints and are
unchanged by a level shift; the one sensitive number is the category
penetration peak, which reads 15.4\% with the step and 14.4\% without
it.

\begin{figure}[t]
  \centering
  \includegraphics[width=\textwidth]{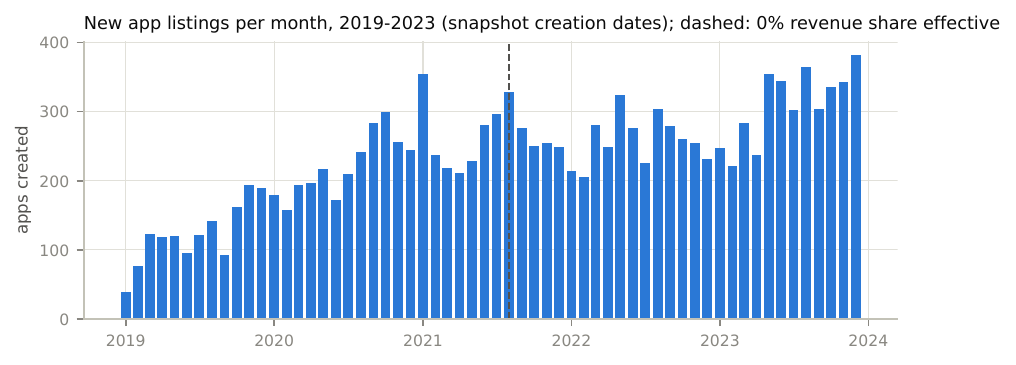}
  \caption{Monthly new application listings, January 2019 to December 2023, from snapshot creation dates over the full application universe. The dashed line marks the revenue-share change (August 2021); the post-2022 rise in entry appears in Figure~\ref{fig:app_timeline}'s annual series.}
  \label{fig:revshare}
\end{figure}

\paragraph{Developer economics: the revenue-share change.}
The reduction of the platform commission to 0\% below \$1M, effective
1 August 2021, produced no detectable change in entry: mean monthly new listings were 262.6 in the twelve months before and 261.2 in the twelve months after (a 0.5\% decrease), on the full application
universe (Figure~\ref{fig:revshare}). The large increase in entry
came later, after 2022 (Section~\ref{sec:rq1}). Retention shows no
detectable response either: the monthly exit hazard among tracked
applications averaged 0.3\% in the year before the change and 0.4\%
in the year after, in line with the general rise in exit rates across
later cohorts (Section~\ref{sec:rq5}), and cohort comparisons across
the boundary are not informative because tracking onboarded few newly
created applications in that period.

\paragraph{Entry without detectable effect: Shopify Forms.}
The launch of Shopify Forms (November 2022) into
email capture shows no detectable incumbent response: median weekly
growth of the leading incumbents was unchanged or improved across the
26-week windows around the launch, so this event contributes no
evidence on platform entry either way.

\subsection{RQ5: Survival and Lifecycle}\label{sec:rq5}

\begin{figure}[t]
  \centering
  \includegraphics[width=0.7\textwidth]{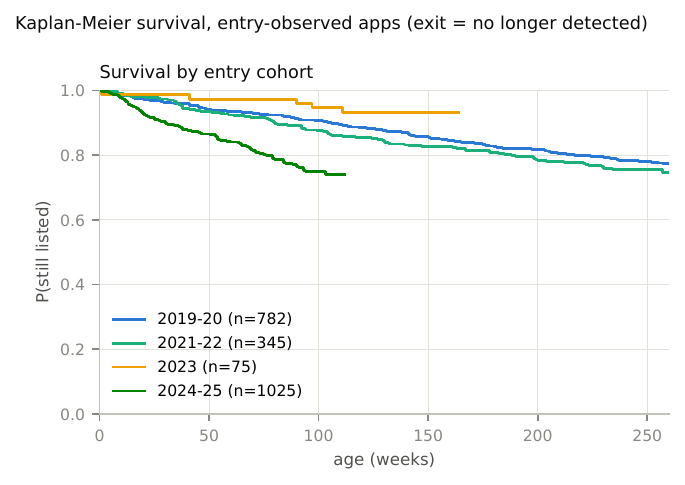}
  \caption{Kaplan--Meier survival of entry-observed applications by
  entry cohort. Each curve shows the share of a cohort still listed
  at each age. The curves cover the 2,227
  entry-observed applications created in 2019 or later; the 20
  created earlier fall outside the cohort bins and are omitted. The
  figure groups entry years as 2019--20, 2021--22, 2023, and
  2024--25, whilst the exit-rate comparison in the text contrasts
  2019--20 with 2023--24.}
  \label{fig:survival}
\end{figure}

Among the 2,247 entry-observed applications, 478 (21.3\%) exited
tracking within the observation window. Tracking exit means the
application is no longer detected on any storefront; validated against
the snapshot, 70.6\% of pre-snapshot tracking exits were also formally
delisted (Inactive), so we report every cohort comparison under both
the tracking and the strict (delisting-confirmed) definition.\footnote{The delisting check is only possible against the September 2025 snapshot: under the definition of Section~\ref{sec:method}, any exit whose last observed week falls on or after the snapshot date counts as strict automatically. Of 867 strict-exit events, 402 (46.4\%) qualify on that clause alone. These unchecked events concentrate in the recent cohorts of the 78-week gradient below (17 of the 2023--24 cohort's 34 events, against none in the two older cohorts); excluding them lowers the 2023--24 strict rate from 7.3\% to about 3.7\%, so the flat gradient, and the conclusion drawn from it, survives the stricter reading.} Exit is
fast and small-scale: the median exiting application peaked at 8
installations 15 weeks after launch and disappeared after 68 weeks,
and 56.5\% of exiting applications never exceeded 10 installations.
Later cohorts lose detectable market presence faster at equal age
(Figure~\ref{fig:survival}): 78-week exit rates rise from 7.5\%
(2019--20 cohorts) to 11.2\% (2023--24), and the gradient persists,
indeed steepens, when onboarding selection is equalised by restricting
to applications first observed with at most 10 installations (9.7\% to
14.1\%). Under the strict definition the gradient flattens (6.4\% to
7.3\%): recent entrants are not delisted faster, they lose their
installed base faster whilst remaining listed.

The Cox model (Table~\ref{tab:cox}) estimates how each entry
characteristic multiplies the risk of exit at any given age; a hazard
ratio below one means lower exit risk. All traction covariates are
measured strictly within the first 26 weeks of life, so that
longer-lived applications do not mechanically score better on them,
and Schoenfeld residual tests show no violation of the proportional
hazards assumption, that is, no evidence that a covariate's effect
changes over an application's life ($p > 0.01$ for every covariate). A freemium tier is associated with a 37\% lower
exit hazard, and receiving any app-store review within the first 26
weeks with a 67\% lower hazard. Each later entry year raises the
tracking-exit hazard by 28\%, with the strict-definition caveat
above. Higher concentration at entry is associated with a lower
hazard in this model, but the association is not robust to
conditioning on early traction (Section~\ref{sec:rq6}), so we treat
it as selection-driven.

\begin{table}[t]
  \centering
  \caption{Cox proportional hazards model of application exit
  (entry-observed applications; $n=1905$, 414 events; early-traction
  covariates measured within the first 26 weeks of life; concordance
  0.713, the probability that the model correctly orders the exit
  times of a random pair of applications).}
  \label{tab:cox}
  \begin{tabular}{lrrrr}
    \toprule
    Covariate & Coef. & Hazard ratio & SE & $p$ \\
    \midrule
    HHI at entry              & $-0.666$ & 0.514 & 0.335 & 0.047 \\
    Freemium                  & $-0.468$ & 0.626 & 0.101 & $<0.001$ \\
    Any review in first 26w   & $-1.093$ & 0.335 & 0.110 & $<0.001$ \\
    Entry year (from 2019)    & $0.246$  & 1.279 & 0.029 & $<0.001$ \\
    \bottomrule
  \end{tabular}
\end{table}

\subsection{RQ6: Early-Warning Prediction of Exit}\label{sec:rq6}

\begin{figure}[t]
  \centering
  \includegraphics[width=\textwidth]{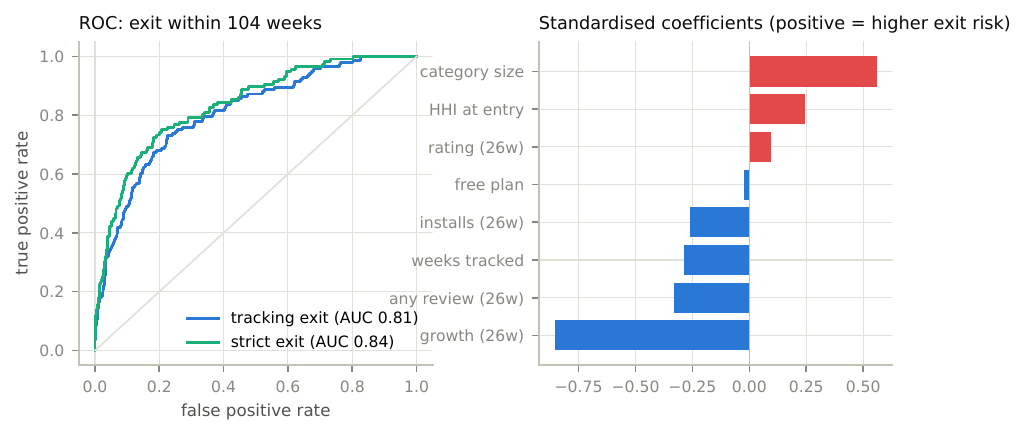}
  \caption{Early-warning prediction of exit within 104 weeks from
  data observable in the first 26 weeks. Left: cross-validated ROC
  curves for both exit definitions. Right: standardised logistic
  coefficients (tracking exit).}
  \label{fig:earlywarning}
\end{figure}

Exit is frequent and foreseeable. A logistic model on eight
features computed strictly within an application's first 26 weeks
(installations reached and added, whether any app-store review was
received, rating, pricing model, weeks tracked, and category size and
concentration at entry) predicts exit within 104 weeks of creation
with a five-fold cross-validated AUC of 0.81 for tracking exit and
0.84 for the strict delisting-confirmed definition
(Figure~\ref{fig:earlywarning}, left; $n = 1{,}309$ entry-observed
applications with full follow-up).

The coefficients read as a compact account of early failure
(Figure~\ref{fig:earlywarning}, right). Installation growth in the
first half-year is by far the strongest protective signal, followed by
receiving any app-store review; absolute installations add little
beyond growth. Entering a larger category is the strongest risk
factor, consistent with crowded functional areas producing most silent
failures, and concentration at entry turns into a risk factor once
early traction is controlled, which is why we read its protective
association in the Cox model as entrant selection instead of shelter.
The small positive coefficient on the early rating does not mean that
good ratings harm survival: it is a conditional effect, measured among
applications with the same growth, review status, and installations,
and applications without a rating receive the median value by
imputation. Unconditionally the association is protective:
applications rated within their first 26 weeks exit within two years
at 7.9\% against 19.7\% for unrated ones, and rated applications at
or above the median rating exit at 5.5\% against 10.8\% below it.
For developers, the practical content is that the first six months of
publicly observable data carry most of the information about two-year
survival: a listing without installation growth and without a first
review by week 26 is already in the highest-risk group, at a point where repositioning remains feasible.

\section{Discussion}\label{sec:discussion}

\paragraph{Winner-take-all, observed in motion.}
The original study argued from a single cross-section that
winner-take-all dynamics do not uniformly characterise the ecosystem.
The panel strengthens that conclusion and adds a direction of travel:
53 of 68 categories with sufficient history de-concentrated significantly over 2019--2026, 13 concentrated, and two showed no significant trend. The two most strongly
concentrating large categories are telling cases: chat concentrated
because the platform owner entered it, and social proof concentrated
around a single dominant application. Product reviews is the more
instructive case: over the panel as a whole it de-concentrated as
the platform application's base dissolved, but from its trough after
the exit concentration rose again as the incumbent leader absorbed
what demand reappeared. In
this ecosystem, sustained concentration appears to require either
platform involvement or an unusually strong leader; left alone,
growing categories drift towards fragmentation. This refines the
original inverse size-concentration finding: growth enables
competitive diversity, and the most prominent exceptions are
governance and dominant-leader stories.

\paragraph{Governance as a market force.}
The event studies show that the platform owner changes competition
far more by building and retiring its own applications than by
changing what it charges developers, and that building and retiring
do not have mirror-image effects. Entry into chat reversed more than
two years of de-concentration
through the platform application's own share capture. Exit from
product reviews did not mirror the effect: at most a third, and
plausibly closer to a sixth, of the released installed base reappeared
as competitor adoption within a year, and the category leader's growth
actually slowed across the event. In the two events observed here, platform entry compressed the space competitors occupy, whilst platform exit did not return it: most of the free default application's installed base did not reappear as competitor adoption. The market the platform leaves is smaller than the one it
entered. By contrast, the revenue-share change, a direct financial transfer to developers, left entry unchanged and retention without detectable response. For marketplace
orchestration this implies that build-and-retire decisions restructure
categories in ways commission levels do not, and that the demand a
platform application appears to hold overstates what any successor can
inherit.

\paragraph{First-mover theory, refined.}
The rolling analysis shows the late-mover advantage is persistent
across seven years, which removes the main temporal-validity concern
of the original study, and the leadership record shows the advantage
reaches the top of categories: the leading application changed in
92\% of categories over the period. The survival analysis, however,
adds a necessary qualification: later cohorts lose detectable market
presence faster at equal age, a gradient that survives the
onboarding-selection check but flattens under the strict delisting
definition, meaning recent entrants are not removed from the store
faster, they fall to zero visible adoption faster whilst remaining
listed. Late entry is therefore a higher-variance strategy that does
not dominate early entry. The applications that enter
late and survive grow faster than incumbents at the same age, whilst a
growing share of their cohort goes silent within eighteen months.
Recency outweighs incumbency for the survivors but not for the cohort.
The exceptions on the incumbent side are equally systematic: the 30\%
of early movers that keep growing are distinguished by leadership,
review volume, ratings, and a free tier, and share the same median
age as the declining majority
(Section~\ref{sec:rq3}), so what looks like an advantage of entering
late is better read as an advantage of being currently good, which
recent entrants and well-run incumbents both hold.

\paragraph{Concentration at entry, reconsidered.}
The Cox model associates higher concentration at entry with a lower
exit hazard, apparently contradicting the original advice that
competitive categories offer better prospects. The early-warning model
resolves the tension: once early traction is held constant,
concentration at entry becomes a risk factor, and category size is the
strongest risk factor overall. The protective association in the
hazard model therefore reflects who chooses to enter concentrated
categories, fewer and more deliberate entrants, and no
shelter from the structure itself, whilst fragmented categories attract
crowds of similar entrants competing at the bottom of the market,
where most silent failures occur. Category selection involves a
trade-off between room to grow and crowding at entry; establishing the
positioning mechanism requires data our sources do not contain.

\paragraph{Failure is predictable, and that changes its meaning.}
That a logistic model on six months of publicly observable data
anticipates two-year exit with AUC above 0.8 says the marketplace's
high failure rate is not noise: most applications that will fail are
already distinguishable from those that will not within their first
half-year. Because the signal is public, the market's verdict on each
application is legible almost immediately, and developers who read it
can reposition early or exit cheaply. Automated monitoring of early
adoption signals is a realistic decision-support instrument for
developers and platform operators alike, and our replication package
contains the code to estimate the model behind one.

\paragraph{Practical implications.}
For developers, the results sharpen the original guidance. Entry
timing matters less than entry conditions: installation growth and a
first review within six months are the strongest observable correlates
of survival, a free tier helps, and proximity to the platform's own
roadmap is a first-order risk, as incumbents in chat learned; but the
deprecation shows that even a platform withdrawal releases less demand
than its installed base suggests. For platform operators, decisions
about what to build and what to retire restructure categories
within quarters, whilst commission changes appear to move neither
entry nor retention much; governance deserves the same scrutiny as
pricing. For
merchants, the prevalence of silent failures underlines the value of
periodically re-evaluating installed applications, particularly after
platform announcements.

\section{Threats to Validity}\label{sec:threats}

\paragraph{Revenue measurement.}
Store revenue estimates represent merchant e-commerce sales rather than app subscription revenue. The \$706 billion figure contextualises merchant base scale but not developer revenue.

\paragraph{Construct validity.}
Install counts combine active and abandoned usage, do not capture monetisation, and mechanically advantage older applications. We mitigated this by restricting to active apps with installs, using relative metrics (market share within categories), and modelling temporal effects through velocity and rate decomposition in RQ3. Store Leads tracks net changes; negative recent values indicate genuine declines in detected installations. Creation date proxies entry timing, conflating age with launch sequence, but remains the best available temporal measure.
In the panel, exit is defined as disappearance from tracking, which
reflects delisting or loss of any detectable storefront presence; both
indicate market exit in the sense relevant here, but the measure
cannot distinguish them. Detection itself can also lapse: Store Leads
recognises an application through its storefront code, so when an
application changes that code its installations can temporarily
disappear from tracking and return once the new signature is
attributed, which in the raw series looks like a decline followed by
a recovery. Two features of the design limit the damage to the exit
results: an application counts as exited only if it stays undetected
to the end of the panel, so any lapse followed by a reappearance is
not an exit, and 70.6\% of pre-snapshot tracking exits are confirmed
delisted in the snapshot, with every cohort comparison repeated under
that strict definition. Declines within a surviving application's
series remain exposed to this artefact, so for the event studies,
where specific declines carry the conclusions, we screened every
series for temporary drops that later recover fully, the signature of
a lapse (a genuine decline does not reverse). No such episode is positioned to
bias the reported numbers, and the one episode material to the deprecation study's windows works in the conservative direction (Section~\ref{sec:rq4}). As an independent activity signal, Judge.me's weekly review count grew by more than 15,000 during the post-shutdown year (May 2024 to May 2025), confirming that its measured deceleration is
slower growth of an active application and not lost detection.
Review growth is only a partial signal, however, because merchant turnover is high enough that an application can gain reviews whilst losing detected installations on net; the interpretation of individual
decline episodes outside the event windows therefore remains an open
validation question.

\paragraph{Internal validity.}
We required categories with at least 20 active apps and analysed only applications aged 180+ days in RQ3, avoiding small-market noise and launch volatility. Early and late movers (first and fourth quartiles by creation date) maximise contrast whilst retaining sample size. Survivorship bias would bias against observing early mover decline; the consistent pattern across categories suggests genuine dynamics rather than artefacts.
The event studies of RQ4 are descriptive comparisons around externally
dated events; they are not causal estimates with matched controls.
Pre-event trends are shown in the figures, redistribution nets out
each competitor's own pre-trend, and the incumbent-growth contrast is
placed against all categories over the same windows by permutation,
which is exactly why we report the chat incumbent slowdown as
suggestive (32nd percentile); the Forms launch shows no detectable
incumbent response and contributes no evidence either way. Confounding
period effects cannot be fully excluded. The association between
concentration at entry and survival is observational, and its sign
reversal under early-traction conditioning indicates entrant
selection.

\paragraph{Panel coverage and censoring.}
The panel tracks 779 applications in 2019 growing to 6,307 in 2026, a
subset of the store skewed towards established applications, and
first appearance in the panel reflects tracking onboarding. We
mitigate this through the snapshot linkage (94.3\% of panel
applications), by restricting cohort and survival analyses to
entry-observed applications, and by computing entry rates around the
revenue-share change from full-universe snapshot creation dates.
Results on panel trajectories are conditional on the tracked
population. Snapshot creation dates can reflect re-listing, which we
guard against with minimum-cell rules in cohort curves.

\paragraph{Archival reconstruction.}
The historical covariates inherit three constraints from their
construction. Archive coverage is uneven: an application enters the
reconstruction only if the Wayback Machine captured its listing, and
semi-annual sampling can miss short-lived price changes; consistent
with industry practice, we assume repricing is infrequent. The
carry-forward join assumes characteristics persist between snapshots;
because every covariate retains its source snapshot date, sensitivity
analyses can be restricted to observations near a snapshot. Finally,
the intersection of the two longitudinal sources (7,708 applications)
is smaller than either source alone and contains only applications
with measurable storefront adoption, so the panel represents the
commercially active segment of the roughly 26,700 applications ever
archived.

\paragraph{External validity.}
Results derive from Shopify (September 2025), which targets SMEs, uses subscription pricing, and curates its store. RQ1 scale metrics likely transfer to other B2B platforms. RQ2 concentration patterns may vary by governance and maturity, though the inverse category size-concentration relationship challenges universal winner-take-all assumptions. RQ3 obsolescence patterns may extend to ecosystems with feasible switching but not high lock-in platforms.
Governance-event findings concern one platform's actions and
generalise as hypotheses only.

\paragraph{Temporal validity.}
Growth metrics in the original study rely on a single 90-day window (July-September 2025), precluding temporal stability assessment. Patterns could reflect period-specific conditions. The systematic nature across 50 categories suggests robustness, though longitudinal data would confirm generalizability.
The panel provides exactly this confirmation: the entry-timing result
is re-estimated at 26 points across seven years and is stable
(Section~\ref{sec:rq123}). The residual constraint is that the panel
ends in March 2026, so post-2025 cohort outcomes are observed only
partially.

\paragraph{Data quality.}
Weekly scraping misses rapid changes. The snapshot's cumulative install field cannot decrease; the panel's detected installations can and do, and they are the measure the exit and event analyses use. Incomplete linking affects demand-side analysis but not concentration or growth measures. Preprocessing reduced inconsistencies. Revenue estimates use traffic models, introducing absolute error but preserving relative patterns.
One three-week scraping anomaly on a single application was identified and documented in the replication package; the application's detected series does not recover afterwards, and because it belongs to no event-study category, no event window's series includes it.

\section{Conclusion}\label{sec:conclusion}

This article extended a cross-sectional study of the Shopify app
ecosystem with a weekly panel spanning February 2019 to March 2026.
The substantive original findings replicate from raw data, are robust to an inclusion difference in the entry-timing sample whose source Section~\ref{sec:rq123} identifies, and hold in time: most
categories de-concentrated as they grew, and the late-mover advantage
observed in September 2025 is the norm across seven years and not an
artefact of one measurement window.

The longitudinal evidence adds three results that a snapshot could
not provide. First, platform governance is a stronger structural
force than developer economics, and it is asymmetric: Shopify's entry
into chat re-concentrated the category through its own share capture,
its exit from product reviews released an installed base of which at
most a third reappeared as competitor adoption within a year, and its revenue-share reduction left entry unchanged and retention without detectable response. In the events observed, platform entry compressed the space competitors occupy, whilst most of the installed base released by platform exit did not reappear as competitor adoption. Second, most failure is silent and early: the
median exiting application peaked at eight installations and
disappeared within sixteen months, later cohorts fall to zero visible
adoption faster at equal age, and yet category leadership changed
hands in 92\% of categories, so the contestability documented in the
original study coexists with rising infant mortality. Third, that
failure is largely foreseeable: six months of publicly observable
data predict two-year exit with cross-validated AUC above 0.8, driven
by early installation growth and the first app-store review.

Together the results refine the original conclusion. App markets on
Shopify remain contestable, but who benefits from that contestability appears to depend less on entry timing than on governance events and entry conditions such as freemium pricing and early review traction. Future
work should test the governance findings on other platforms, explain
why the minority of early movers that resists obsolescence does so,
which requires firm-level information beyond public listings, and
connect adoption trajectories to monetisation, which no
storefront-scraping source can observe.

\section*{Declarations}
\paragraph{Competing interests.}
The first two authors are employed by Judge.me. All adoption data
comes from an independent third-party provider (Store Leads), the
panel analyses are fully scripted in the replication package, and the event
dates derive from public announcements.
\paragraph{Data availability.}
A replication package accompanies this article and is available at
\href{https://github.com/giuseppedestefanis/scale-concentration-entrytime-shopify-ecosystem}{this link}.\footnote{\url{https://github.com/giuseppedestefanis/scale-concentration-entrytime-shopify-ecosystem}} It
contains the analysis scripts and figure-generation code for the
panel results, the archival reconstruction pipeline of
Section~\ref{sec:dataset} (discovery, acquisition, extraction, and
validation), and the data that pipeline produced from public
sources: the extracted pricing, category, and listing-metadata
records. Store Leads data
cannot be redistributed under its commercial licence. This covers
the September 2025 snapshot exports, the weekly installation
counts, and the weekly review counts, so the merged application-week
panel and every file derived
from them are excluded from the package. The package documents the
exact extracts required, and its input-validation script checks a
purchased extract before the pipeline runs, so the full analysis is
reproducible for readers who obtain the same data from the
provider.

\end{document}